\documentclass[a4paper,11pt]{article}
\pdfoutput=1 

\usepackage{jcappub} 

\usepackage[T1]{fontenc} 
\usepackage{placeins}
\usepackage{comment}
\usepackage{graphicx}
\usepackage{subfig}
\usepackage{cleveref}
\usepackage[a4paper, total={6in, 9in}]{geometry}

\title{\boldmath Dark Matter Production in an Early Matter Dominated Era}

\author[a]{Manuel Drees,}
\author[a,1]{Fazlollah Hajkarim,\note{Corresponding author.}}

\affiliation[a]{Bethe Center for Theoretical Physics and Physikalisches
  Institut, Universit\"at Bonn,\\Nussallee~12, D-53115 Bonn,
  Germany}

\emailAdd{drees@th.physik.uni-bonn.de}
\emailAdd{hajkarim@th.physik.uni-bonn.de}

\abstract{We investigate dark matter (DM) production in an early
  matter dominated era where a heavy long--lived particle decays to
  radiation and DM. In addition to DM annihilation into and thermal DM
  production from radiation, we include direct DM production from the
  decay of the long--lived particle. In contrast to earlier treatments
  the temperature dependence of the number of degrees of freedom $g_*$
  in the Standard Model (SM) plasma is treated carefully. Besides the
  well--known cases of thermal hot and cold DM, additional regions of
  parameter space with the approximately correct DM relic density
  appear. In some of these regions the temperature dependence of $g_*$
  can change the final DM density by several hundred
  percent. Furthermore, we analyze the effect of allowing nonvanishing
  initial abundances for radiation and DM. We find an upper bound on
  the mass of the long--lived particle if the DM annihilation cross
  section is below that corresponding to thermal WIMP (Weakly
  Interactive Massive Particle) DM in standard cosmology. }

\begin{document}
\def\gsim{\:\raisebox{-0.5ex}{$\stackrel{\textstyle>}{\sim}$}\:}
\def\lsim{\:\raisebox{-0.5ex}{$\stackrel{\textstyle<}{\sim}$}\:}

\maketitle
\flushbottom

\section{Introduction}
\label{intro}

The nature of Dark Matter (DM) has been a mystery for several
decades. Most proposals for its explanation \cite{Bertone:2016nfn}
require new particle physics, since astrophysical and cosmological
observations imply that DM consists of cold particles (which were
non--relativistic at the onset of structure formation)
\cite{Bertone:2004pz,Baer:2014eja}. The Standard Model (SM) of
particle physics does not contain any such particle, whereas many
extensions of the SM do. 

A widely studied class of DM candidate particles are Weakly
Interacting Massive Particles (WIMPs) \cite{Bertone:2004pz}. They can
be thermally produced in the early universe. This means that at
sufficiently high temperature they were in thermal equilibrium with
the plasma of SM particles. However, as the universe expanded and
hence cooled, their abundance dropped, until the WIMP annihilation
rate equaled the Hubble expansion rate. At this temperature WIMPs
decoupled (``froze out''), meaning their comoving density became
essentially constant. In standard cosmology this mechanism requires a
specific value for the (thermally averaged) WIMP annihilation cross
section in order to reproduce the observed DM density; this cross
section ``happens'' to be close to a typical weak cross section (hence
the name). For example, supersymmetric (SUSY) extensions of the SM
contain WIMP candidates \cite{susy}.

Thermal WIMPs are attractive since they can be searched for in several
different ways. However, neither direct
\cite{Angloher:2015ewa,Agnese:2015nto,Akerib:2016vxi,Tan:2016zwf} nor
indirect \cite{Ackermann:2015zua,Fermi-LAT:2016uux,Ahnen:2016qkx} WIMP
searches have found any signal as yet, and collider searches for
particles not contained in the SM have also only yielded (a large
number of) bounds but no positive evidence. This has led to renewed
interest in extensions of the simple thermal WIMP scenario.

One possibility is to consider a modified expansion history of the
universe. In standard cosmology one assumes that the universe became
radiation dominated after the end of inflation, and stayed that way
down to a temperature of about $1$ eV, at which point it became
dominated by (mostly dark) matter. However, string theory and other
UV--complete theories suggest that there may have been an epoch of
early matter domination after inflation and before Big Bang
Nucleosynthesis (BBN), which must have occurred in a
radiation--dominated era. This early matter domination would have been
due to massive scalar particles with very weak couplings to SM
particles, and hence long lifetimes. In string theory the vacuum
expectation values (VEVs) of such ``moduli'' fields determine the
couplings of the low energy theory
\cite{Easther:2013nga,Arcadi:2011ev}. In fact, already supergravity
theories where supersymmetry is broken in a hidden sector contain
scalar (``Polonyi'' \cite{polonyi}) fields with similar
properties. These scalar fields obtained large values during
inflation, if their mass was smaller than the Hubble scale during
inflation \cite{modprod1,modprod2,modprod3,modprod4,modprod5}. Later these fields started to oscillate
coherently, which corresponds to an ensemble of scalar particles at
rest in the cosmic rest frame.

If these fields dominated the energy density of the universe, their
decay produced a lot of entropy. This would have diluted the density
of all particles that had been produced before. Moreover, these decays
may have occurred so late that they affected the (largely) successful
predictions of standard BBN; this is the cosmological moduli (or
Polonyi) problem
\cite{Polnarev:1982,Coughlan:1983ci,deCarlos:1993wie,Banks:1993en,Ellis:1986zt}.
The detailed analysis \cite{Kawasaki:2000en,Hannestad:2004px} showed that the reheat
temperature, i.e. the highest temperature of the radiation--dominated
epoch, must have been at least $\sim 4$~MeV in order not to jeopardize
the success of standard BBN; this bound has more recently been
confirmed in \cite{deSalas:2015glj}. We will see below that this
requires the scalar mass to be well above $10$ TeV. Such a large mass
is problematic if this is also the scale of visible--sector
superparticle masses, and one wishes to use supersymmetry to solve the
hierarchy problem. (However, large superparticle masses are still
acceptable for ``split'' Supersymmetry \cite{split1,split2}.) Any realistic
model assuming that DM is produced during an early epoch of matter
domination by a heavy scalar \cite{Acharya:2009zt,Gelmini:2006pw} has to
respect this bound.

Early studies of the non--thermal production of DM particles focused
on the reheating era at the end of inflation
\cite{Chung:1998rq,Giudice:2000ex,Allahverdi:2002nb,Pallis:2004yy};
more recently, this has been analyzed in \cite{Dev:2013yza}.

In supersymmetric or superstring theories decays of the gravitino can
also cause problems with BBN (gravitino problem). Some models that
solve this problem predict a period of modulus domination
\cite{Allahverdi:2013noa}. In other scenarios gravitino decays do not
overproduce DM \cite{Nakamura:2006uc}, or the gravitino is itself a
stable DM candidate \cite{Asaka:2006bv,Feng:2004mt}. In some
supersymmetric scenarios moduli can decay to gravitinos; in this case
the gravitino mass should be high enough to prevent its decay at BBN
time (moduli-induced gravitino problem)
\cite{Nakamura:2006uc,Endo:2006zj,Ellis:1986zt}. On the other hand, if
the gravitino mass is larger than that of the moduli
\cite{Blumenhagen:2009gk,Aparicio:2014wxa,Allahverdi:2013noa} a
solution of the moduli problem automatically solves the gravitino
problem as well. In our analysis we will ignore the gravitino,
implicitly assuming that one of these solutions is at work.

Generally there are two thermal DM production mechanisms. The
freeze--out (FO) scenario for WIMP DM has been described above; here
the relevant dynamics occurs around the freeze--out temperature, which
is typically $5\%$ of the WIMP mass. In contrast, in the freeze--in
(FI) scenario dark matter is produced at temperatures above the mass
of the DM particle, e.g.  by the decay of a heavier particle. Here the
DM annihilation cross section is so small that DM annihilation is
negligible \cite{Giudice:2000ex,Hall:2009bx}. These two general
processes can also happen in an early matter dominated era, since the
decays of the heavy scalar will generate a (subdominant) radiation
component. However, other possibilities exist for DM production during
such an epoch. These have been explored in \cite{Kane:2015qea}, which
in addition considered the decay of a WIMP--like visible sector
particle into a lighter DM particle residing in a hidden sector.

The present analysis is based on \cite{Kane:2015qea}. However, we
assume that DM resides in the visible sector, and do not include a
dark radiation component (which is quite strongly constrained by
recent cosmological data \cite{Ade:2015xua}). We instead focus on a more
careful description of the thermal history of the universe. This
includes effects due to the temperature dependence of the effective
number of relativistic degrees of freedom $g_*$ (and of the analogous
quantity defined via the entropy density rather than via the energy
density of the radiation). Here we use the results of
\cite{Drees:2015exa}, which assumes free electroweak gauge and Higgs
bosons, as appropriate for a smooth crossover electroweak transition
\cite{Kajantie:1996mn,Csikor:1998eu,Fodor:1999at}. Moreover, it uses
results from lattice QCD around the QCD transition temperature,
matched to a hadron resonance gas at lower temperatures
\cite{Bazavov:2014pvz,Huovinen:2009yb}. Finally, at MeV temperatures
neutrino decoupling is treated using results from
\cite{Lesgourgues:2012uu}. Moreover, we also consider scenarios with
non-vanishing radiation and DM content at the onset of the early matter
dominated epoch.
 
The remainder of this article is organized as follows. In the next
Section we describe the general formalism for computing the DM relic
density in a cosmology with an early matter dominated epoch. Here we
largely follow ref.\cite{Kane:2015qea}, but with improved treatment
of the radiation component. In Sec.~\ref{dmprod} we map out regions
of parameter space giving the correct DM relic density assuming initially
vanishing DM and radiation content, with special emphasis on the effect
of the temperature dependence of $g_*$. In Sec.~\ref{initdarkrad} we
then allow non--vanishing initial radiation and DM density, before
concluding in Sec.~\ref{conclusion}.

\section{The General Framework for Non-Thermal Dark Matter Production}
\label{nonthermaldm}

In this Section we describe the calculational framework for computing
the DM relic density in a scenario with an early epoch of matter
domination. In the first Subsection we define the variables we use,
and the equations that determine their evolution during the early
universe. Our analysis is model--independent in the sense that all
relevant particle physics quantities -- particle masses, the DM
annihilation cross section and the decay width of the heavy scalar
particle -- are treated as free parameters. This discussion is based
on ref.\cite{Kane:2015qea}, but we extend it by correctly treating the
temperature dependence of the number of relativistic degrees of
freedom in the thermal plasma. In the second Subsection we briefly
describe the numerical solution of the evolution equations, and the
relation of the dimensionless quantities introduced in the first
Subsection to the scaled DM relic density.

\subsection{Evolution Equations}
\label{friedmann}

In the standard thermal DM production scenarios (both freeze--in and
freeze--out) one only needs to solve a single evolution equation, namely
the Boltzmann equation for the number density of the DM particles. In
these scenarios all the relevant dynamics happens during the radiation
dominated epoch, so the state of the universe is essentially determined
uniquely by the temperature $T$. In particular, both the Hubble parameter
and the entropy density are functions of $T$ only; moreover, the comoving
entropy density is constant in this case, since the universe evolves
adiabatically.

This is no longer true in the case we are interested in, where the
energy density of the universe was dominated for a while by slowly
decaying scalar $\phi$ particles.\footnote{The spin of the decaying
  particles is not relevant for us. However, the by far best motivated
  particle physics realizations of this mechanism use scalar moduli or
  Polonyi fields, as discussed in the Introduction.} We therefore need
to track three coupled evolution equations: for the DM particles, for
the $\phi$ particles, and for the radiation content.

Following \cite{Giudice:2000ex,Arcadi:2011ev,Kane:2015qea} we
introduce dimensionless quantities in order to describe the evolution
of the universe. To this end all dimensionful quantities are divided
by appropriate powers of the ``reheat temperature'', defined as
\begin{equation} \label{eq:T_RH}
T_{\rm RH} = \sqrt{\Gamma_\phi M_{\rm Pl}} \left( \frac {45} {4 \pi^3 g_*(T_{\rm RH})}
 \right)^{1/4}\,.
\end{equation}
Here $M_{\rm Pl} \simeq 1.22 \times 10^{19}$ GeV is the Planck mass,
$\Gamma_\phi$ is the total decay width of the $\phi$ particles, and
$g_*$ is the number of relativistic degrees of freedom in the thermal
plasma, defined via the energy density of radiation \cite{kt}:
\begin{equation}\label{rho_R}
\rho_{R}(T) = \frac{\pi^2}{30} g_*(T) T^4\,.
\end{equation} 
Note that $T_{\rm RH}$ is a bit of an idealization: it is the
temperature the universe would have if all the energy stored in $\phi$
was instantaneously transformed into radiation at Hubble parameter
$H = \Gamma_\phi$, under the assumption that $\phi$ particles
completely dominated the energy density of the universe before their
decay
\cite{Chung:1998rq,deSalas:2015glj,Kane:2015jia,Pallis:2004yy,Arcadi:2011ev}.
Nevertheless $T_{\rm RH}$ is a good estimate of the highest
temperature of the radiation--dominated epoch that begins after most
$\phi$ particles have decayed. Note, however, that the thermal bath
during moduli domination can be considerably hotter than $T_{\rm RH}$;
we will come back to this point below.

The decay width of moduli (or Polonyi) fields $\phi$ are Planck
suppressed, but the precise coupling strength is model dependent. We
thus write
\begin{equation} \label{decaywidth}
\Gamma_\phi = \alpha \frac{M_\phi^3} {M_{\rm Pl}^2}\,, \ \ 
\alpha=\frac{C}{8 \pi}=~{\rm constant}\,,
\end{equation}
where $M_\phi$ is the mass of $\phi$, and $C$ is a constant whose
value depends on the UV--complete theory. The value of $T_{\rm RH}$ is
obtained by computing $\Gamma_\phi$ (which requires fixing $M_\phi$
and $C$ or $\alpha$), and inserting this into eq.(\ref{eq:T_RH}). Note
that this is an implicit equation, since the right--hand side (rhs)
depends on $T_{\rm RH}$ via $g_*$. We use results from
ref.\cite{Drees:2015exa} to compute the temperature dependence of
$g_*$.

\begin{figure}
  \centering
\includegraphics[width=0.85\textwidth]{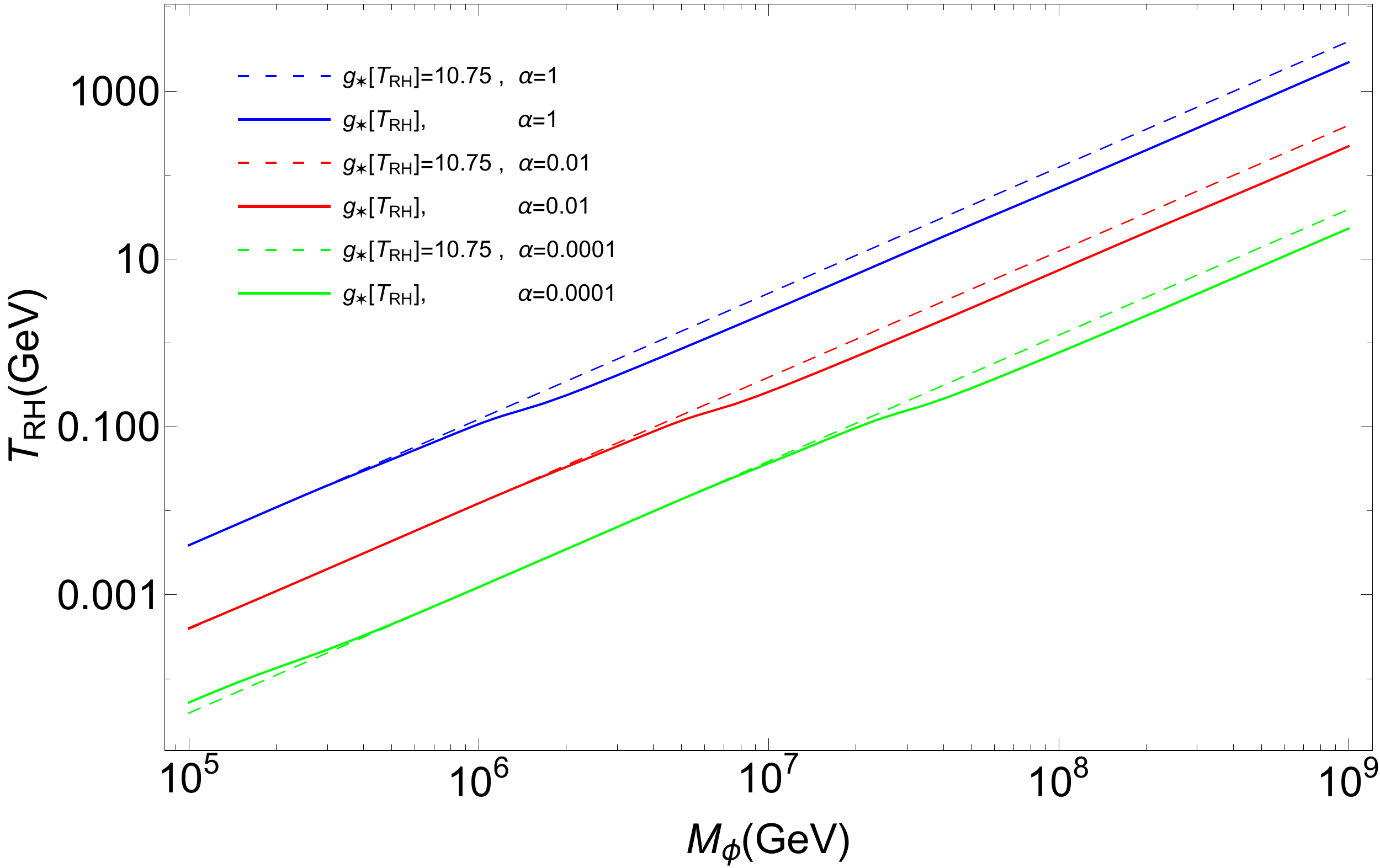}
\\
\caption{Reheating temperature $T_{\rm RH}$ as function of
  the mass $M_\phi$ of the particle whose energy density dominates in
  the early matter dominated epoch, for different coefficients
  $\alpha$ defined in eq.(\ref{decaywidth}). The solid curves have
  been obtained including the temperature dependence of $g_*$ as
  predicted by the SM, whereas the dashed curves are for fixed
  $g_*(T_{RH}) = 10.75$.}
\label{reheatingtemp1}
\end{figure}

As noted above, the success of standard BBN requires
$T_{\rm RH} \gsim 4$~MeV. This implies $M_{\phi} \gsim 100$~TeV for
$\alpha \sim 1$ \cite{Kawasaki:2000en,Hannestad:2004px,deSalas:2015glj}. In
Fig.~\ref{reheatingtemp1} the dependence of the reheating temperature
on the modulus mass according to eq.(\ref{eq:T_RH}) is shown for
different couplings $\alpha$ between $10^{-4}$ and $1$. Here we have
assumed that only SM particles contribute to $g_*$. If some new
particles are found, as predicted e.g. by supersymmetric extensions of
the SM, this figure will definitely change at higher reheating
temperatures, i.e for larger $M_\phi$. For comparison,
Fig.~\ref{reheatingtemp1} also shows results for fixed
$g_*(T_{\rm RH}) = 10.75$, as appropriate for the SM at temperatures
before BBN but after the decoupling of muons. In this plot, with its
logarithmic axes, the differences between the two sets of curves
become apparent only for $T_{\rm RH}$ of order the QCD transition
temperature $T_c \sim 150$~MeV or higher.\footnote{The curves also
  differ slightly for $T_{\rm RH}$ below the electron mass, where the
  actual $g_*(T_{\rm RH})$ is less than $10.75$. However, scenarios
  with such a low reheat temperature will not reproduce standard BBN.}
However, this is by far not the only way in which the temperature
dependence of $g_*$ affects the final result.

Our main interest is the calculation of the Dark Matter relic density,
by deriving and solving the relevant Boltzmann equations. To that end,
we use $T_{\rm RH}$ to define the dimensionless scale parameter
\begin{equation} \label{eq:A}
A \equiv a T_{\rm RH}\,;
\end{equation}
multiplying the scale factor in the Friedmann--Robertson--Walker
metric $a$ with $T_{\rm RH}$ improves the stability of the numerical
solution \cite{Arcadi:2011ev}. This in turn allows us to define
dimensionless co--moving densities for $\phi$ particles, radiation and
DM particles:
\begin{equation} \label{eq:densities}
\Phi \equiv  \frac{\rho_{\phi} A^3}{T_{\rm RH}^4}\,, \,\, R \equiv \rho_R \frac{A^4}
{T_{\rm RH}^4}\,, \,\, X^\prime \equiv n_{X^\prime} \frac{A^3}{T_{\rm RH}^3}\,.
\end{equation}
Here $\rho_\phi = M_\phi n_\phi$ is the energy density stored in
$\phi$ particles, and $n_{X^\prime}$ is the number density of the DM
particles, which we call $X^\prime$ following
ref.\cite{Kane:2015qea}. Note that $R, \, \Phi$ and $X^\prime$
approach 
constants when $\phi$ decays as well as the pair production
and annihilation of $X^\prime$ particles can be neglected. Finally, we
use the comoving densities to define a dimensionless comoving Hubble
parameter:
\begin{equation} \label{eq:hubble_1}
\widetilde{H} \equiv \left(\Phi + \frac{R}{A} + 
\frac{E_{X^\prime}  X^\prime }{T_{\rm RH}} \right)^{1/2}\,.
\end{equation}
Here $E_{X^\prime} \approx (M_{X^\prime}^2 + 3\, {T}^2)^{1/2}$ is the
average energy per $X^\prime$ particle; this approximation is
sufficient for our purposes since the contribution of DM particles to
the total energy density is always subdominant in the epoch we are
interested in.\footnote{This ansatz implicitly assumes that $X^\prime$
  particles are at least in kinetic equilibrium with the SM
  radiation. Note that kinetic equilibrium is much easier to attain
  than full chemical equilibrium.} $\widetilde{H}$ is related to the
usual (dimensionful) Hubble parameter via
\begin{equation} \label{eq:hubble_2}
H= \widetilde{H} T_{\rm RH}^2 A^{-3/2} c_1^{-1/2} M_{\rm Pl}^{-1} \,,
\end{equation}
where we have introduced the constant $c_1 = 3 / (8 \pi)$.

As in ref.\cite{Kane:2015qea} we assume that $\phi$ particles can
decay into $X'$ particles with effective branching ratio
$B_{X^\prime}$; the average energy per $\phi$ decay that goes into DM
particles is then given by $\bar{B} M_\phi$, with
\begin{equation} \label{eq:Bbar}
\bar{B} = \frac {E_{X^\prime} B_{X^\prime}} {M_\phi}\,.
\end{equation}
A fraction $1 - \bar{B}$ of the $\phi$ energy will then go into SM
particles, i.e. into radiation. In many cases a discrete symmetry
ensures that $X^\prime$ particles can only be produced pairwise. If
$\phi \rightarrow X^\prime X^\prime$ is the dominant $X^\prime$
production mode from $\phi$ decays, then
$B_{X^\prime} = 2 \Gamma(\phi \rightarrow X^\prime X^\prime) /
\Gamma_\phi$.\footnote{$\phi$
  particles might decay predominantly into heavy SM particles,
  e.g. top quarks or Higgs bosons, with masses larger than the
  temperature. However, these heavier SM particles will then decay
  almost immediately into light SM particles, i.e.  into
  radiation. $\phi$ particles could also decay into some partners of
  $X^\prime$, e.g. a pair of gluinos in supersymmetric models, which
  then decay almost immediately into $X^\prime$ plus radiation. All
  these cases are covered by this formalism.} Note that we will mostly
be interested in scenarios where $B_{X^\prime} \ll 1$, i.e.
$\bar B \ll 1$; the exact expression for $E_{X^\prime}$ is then again
not important.

In the following analysis we assume that the decay products of $\phi$
thermalize immediately, i.e. radiation always refers to a relativistic
plasma in full thermal equilibrium, again following
ref.\cite{Kane:2015qea}. This is an idealization. If $\phi$ particles
decay into two body final states, these final state particles will
initially have energy $M_\phi/2$, which can be much higher than $T$.
These energetic SM particles could produce DM particles before they
thermalize. This has been analyzed in \cite{allah2}, where it was
shown that this source of $X^\prime$ particles can be significant if
$M_{X^\prime}$ is relatively close to $M_\phi$. We are mostly
interested in $M_{X^\prime} \ll M_\phi$, in which case the
approximation of instantaneous thermalization of the $\phi$ decay
products should be applicable. Recently, it has been claimed that considering 
the details of thermalization process can change the maximum temperature of the universe 
and affect the process of DM production during and after early matter domination 
\cite{Harigaya:2013vwa,Harigaya:2014waa,Mukaida:2015ria}. We postpone to consider these details to future studies.

The effective number of degrees of freedom $g_*$, defined via the
radiation energy density as in eq.(\ref{rho_R}), depends on the
temperature, since only particles with mass of order of or less than
the temperature contribute significantly \cite{kt}. In the SM the
temperature dependence is rather mild for $T~\gsim~1$ GeV. As a first
approximation one can therefore ignore terms proportional to the
derivative $d g_*/dT$. This allows to derive an evolution equation for
$R$ from energy conservation. The set of equations one needs to solve
is then \cite{Kane:2015qea}:
\begin{eqnarray} \label{eq:boltzmann}
\notag \widetilde{H} \frac{d\Phi}{dA} &=& -\,c_{\rho}^{1/2}\,A^{1/2} \Phi , 
\notag  \\
\widetilde{H} \frac{d R}{dA} &=& c_{\rho}^{1/2}\,A^{3/2} (1 - \bar{B} )  \Phi 
+ c_1^{1/2}\,M_{\rm pl}\,\left[ \frac{2 E_{X^\prime} \langle\sigma v \rangle'}
{A^{3/2}} \left({X^\prime}^2 - {X^\prime_{\rm EQ}}^2 \right) \right],  \\
\widetilde{H} \frac{d X^\prime}{d A } &=& \frac{c_\rho^{1/2} T_{\rm RH} B_{X^\prime}}
{M_{\phi}} A^{1/2} \Phi + c_1^{1/2}\, M_{\rm pl} T_{\rm RH} A^{-5/2} \, 
\langle \sigma v \rangle' \left( {X^\prime_{\rm EQ}}^2 - {X^\prime}^2 \right).
\notag
\end{eqnarray}
Here $c_1$ is as in eq.(\ref{eq:hubble_2}), and we have defined a
second constant $c_{\rho} = \frac{\pi^2 g_*(T_{\rm RH})} {30}$. Finally,
the scaled $X'$ equilibrium number density $X^\prime_{\rm EQ}$ is given by
\begin{equation} \label{eq:X_EQ}
X^\prime_{\rm EQ} \equiv \left( \frac{A} {T_{\rm RH}} \right)^3 
\frac{g_{X^\prime} T {M_{X^\prime}}^2} {2 \pi^2} K_2 \left( 
\frac{M_{X^\prime}} {T} \right)  \rightarrow {\small \left\{ \begin{array}{l}
\left( \frac{A} {T_{\rm RH}} \right)^3 \frac{ \tilde{g} \,\zeta(3) T^3} {\pi^2}
\hspace*{3.4cm} {\rm if} \ T \gg M_{X^\prime} \\
\left( \frac{A} {T_{\rm RH}} \right)^3 g_{X^\prime} \left( 
\frac {M_{X^\prime} T} {2 \pi} \right )^\frac{3}{2} \exp(-M_{X^\prime}/T) \ \ 
{\rm if} \ T \ll M_{X^\prime} 
\end{array} \right.}
\end{equation}
Here $g_{X^\prime}$ counts the internal degrees of freedom of
$X^\prime$, $\tilde{g} = g_{X^\prime}\, (3 g_{X^\prime}/4)$ for
bosonic (fermionic) $X^\prime$, and $K_2$ is the modified Bessel
function of second kind. In our numerical calculations we assume
$g_{X^\prime}=2$, as appropriate for a spin$-1/2$ Majorana
(self--conjugate) fermion.

The first eq.(\ref{eq:boltzmann}) describes $\phi$ decays.
Unfortunately it is not entirely straightforward to see in this
formalism that $\Phi$ becomes constant when $\phi$ decays can be
neglected. Eq.(\ref{eq:T_RH}) shows that $T_{\rm RH} \rightarrow 0$ as
$\Gamma_\phi \rightarrow 0$, so eqs.(\ref{eq:densities}) become
ill--defined in this case. Note, however, that initially $\Phi$, and
hence $\widetilde H$, are much bigger than unity if moduli are to
dominate the universe for an extended epoch. Initially $d \Phi / dA$
is thus much less than $\Phi$. One can show that $A d \Phi / d A$
becomes of order $\Phi_I$ only when $H \simeq \Gamma_\phi$.

In the second eq.(\ref{eq:boltzmann}) we recognize a positive
contribution to the rhs proportional to $\Phi$ stemming from $\phi$
decays, and a contribution describing the pair production from and
annihilation of $X'$ particles into radiation. A similar term appears
on the rhs of the third equation with opposite sign; this third
equation also features a positive contribution from direct
$\phi \rightarrow X'$ decays.
 
In order to follow the evolution of the universe more accurately we
must consider the precise evolution of $g_*$ (and related quantities)
in the thermal bath. As shown in ref.\cite{kt} the evolution of the
radiation component is then more easily described via the entropy
density, which is given by
\begin{equation} \label{eq:s_R}
s_R (T) = \frac{\rho_R(T)+p_R(T)}{T} = \frac{2\pi^2}{45} h_*(T) T^3\,. 
\end{equation}
The second equation defines $h_*(T)$, which is another measure of the
effective number of relativistic degrees of freedom; in the SM, $g_*(T) = 
h_*(T)$ before neutrinos decouple at MeV temperatures. In standard cosmology,
the comoving entropy density is constant after the end of inflation. However,
in the scenario considered here $\phi$ decays lead to entropy production.
This is described by the evolution equation
\begin{equation} \label{entropydensity}
\frac {d\, s_R} {dt} + 3 Hs_R = \frac{1}{T} \big[ (1 - \bar{B}) \Gamma_{\phi }
 \rho_{\phi} + 2 E_{X^\prime} \langle \sigma v \rangle^\prime 
\left( n_{X^\prime}^2 - {n_{X^\prime,\rm EQ}}^2\right) \big].
\end{equation}
The factor $(1- \bar{B})$ in the first term of the rhs of
eq.(\ref{entropydensity}) should not be there if $T > M_{X^\prime}$; recall,
however, that $\bar{B} \ll 1$ in cases of interest, so that we make only a
small mistake by including this factor. The second term describes entropy
production from out--of--equilibrium annihilation of $X'$ particles; we find 
that this term is always numerically insignificant. [This is true also for the
last term on the rhs of the second eq.(\ref{eq:boltzmann})]. 

This leads to the following equation describing the evolution of the
temperature:
\begin{eqnarray} \label{temp2}
\nonumber \frac{dT}{dA} = \left( 1 + \frac{T}{3 h_*} \frac{dh_*}{dT} 
\right)^{-1}  &\Bigg[& -\frac{T}{A} + \frac {15 T_{\rm RH}^6} 
{2 \pi^2 c_1^{1/2} M_{\rm Pl} H T^3 h_* A^{\frac{11}{2}} } 
\Bigg( c_{\rho}^{1/2}\,A^{3/2} (1 - \bar{B})  \Phi 
\\ 
 && + c_1^{1/2}\, M_{\rm pl}\, \frac{2 E_{X^\prime} \langle \sigma v \rangle'} 
{A^{3/2}} \left({X^\prime}^2 - {X^\prime_{\rm EQ}}^2 \right)\Bigg)
 \Bigg]\,.
\end{eqnarray}
Note that the rhs of eq.(\ref{temp2}) depends both on $h_*(T)$ and
(via $c_\rho$) on $g_*(T)$. We will use the results of
\cite{Drees:2015exa} for them. This equation replaces the second
eq.(\ref{eq:boltzmann}); the first and third of these equations remain
unchanged. We also need eq.(\ref{rho_R}) to compute the radiation
density from $T$, and eqs.(\ref{eq:hubble_1}) and (\ref{eq:densities})
to compute the scaled Hubble parameter $\widetilde H$. This is a
closed system of equations.

As mentioned above, in the early epoch of matter domination the
radiation component can be much hotter than $T_{\rm RH}$. If terms
proportional to the derivative of $h_*$ or $g_*$ with respect to
temperature are ignored, the evolution of the temperature for
$H \gg \Gamma_\phi$ can be computed analytically. If initially
$\rho_R = 0$ one finds \cite{Giudice:2000ex}:
\begin{equation} \label{tempa}
T \simeq \left( \frac{8^8}{3^3 5^5} \right)^{1/20} \left( \frac {g_*(T_{\rm max})}
{g_*(T)} \right)^{1/4} {T_{\rm max}} \left( A^{-3/2} - A^{-4} \right)^{1/4}\, .
\end{equation}
The maximum temperature during modulus domination $T_{\rm max}$ is given by 
\cite{Giudice:2000ex}
\begin{equation} \label{tmax}
T_{\rm max} =  \left( \frac {3} {8} \right)^{2/5} \left( \frac {5} {\pi^3}
\right)^{1/8} \left( \frac {g_*(T_{\rm RH})^{1/2}} {g_*(T_{\rm max})} \right)^{1/4} 
(M_{\rm pl} H_I T_{\rm RH}^2)^{1/4}\, .
\end{equation}
Using $H_I = \Phi_I^{1/2} T_{\rm RH}^2 / (c_1^{1/2} M_{\rm Pl})$ we find
$T_{\rm max} \sim T_{\rm RH} \Phi_I^{1/8}$, up to ${\cal O}(1)$
factors, where $\Phi_I$ is the initial co--moving density of $\phi$. The assumption of vanishing initial $\rho_R$ is reasonable
only if the matter dominated epoch lasts sufficiently long to
basically erase all traces of possible earlier radiation dominated
epochs. This requires $H_I \gg \Gamma_\phi$, and hence
$\sqrt{\Phi_I} \gg 1$. Hence even if $T_{\rm RH}$ is well below the
temperature $T_c$ of the QCD transition, frequently $T_{\max} > T_c$, in
which case an accurate treatment of the temperature dependence of $g_*$
and $h_*$ becomes important.

\subsection{Predicted Dark Matter Abundance}
\label{dmrelic}

As in case of standard cosmology, the dark matter annihilation cross
section plays an important role. Here we will not consider specific
particle physics candidates for $X^{\prime}$. Instead we use the
standard parametrization,
\begin{equation} \label{eq:sigma}
 ( \sigma v )^\prime=a + b v^2\,,
\end{equation}
which is applicable for particles whose final relic density is set at
temperature $T < M_{X^\prime}$. Here $a$ is nonzero only if $X^\prime$
particles can annihilate from an $S-$wave initial state, whereas a
non-vanishing $b$ can be generated also by annihilation from the
$P-$wave. Thermal averaging then gives
\begin{equation} \label{eq:sigav}
\langle \sigma v \rangle^\prime =a +6b \frac {T} { M_{X^\prime} }\,.
\end{equation}

In the following chapters we will present exact numerical solutions of the
evolution equations discussed in the previous subsections. These have
been obtained with the help of {\tt Mathematica}. We found it
convenient to rewrite the equations in terms of the logarithmic
derivative $d / d (\ln A) = A d / d A$, i.e. we multiplied
eqs.(\ref{eq:boltzmann}) and (\ref{temp2}) with $A$. We always use
initial conditions
\begin{equation} \label{initialc1}
A = 1, \,\,\Phi = \Phi_I = \frac{3H_I^2\,M_{\rm pl}^2}{8\pi\,T_{\rm RH}^4}\,;
\end{equation} 
in the next chapter we will follow ref.\cite{Kane:2015qea} in
initially setting
\begin{equation} \label{initialc2}
R_I = X^\prime_I = 0\,.
\end{equation}
The assumption of initial density for radiation and dark matter
particles can be reasonable if $\phi$ particles completely dominate
the universe for some time after inflationary reheating, so that all
dependence on conditions before the $\phi$ dominated epoch is
erased. Recall that in the absence of $\phi$ decays the ratio of
radiation and matter densities scales like $1/A$. A possible initial
radiation component can then become irrelevant if $\tilde A \gg 1$,
where $\tilde A$ is the value of the dimensionless scale factor where
most $\phi$ particles decay. Until this time to good approximation
$H_I = $~constant. The first eq.(\ref{eq:boltzmann}) can then be
solved analytically \cite{Kane:2015qea}:
\begin{equation}
\Phi(A) = \Phi_I \, \exp \left[ -\frac{2}{3} \left( \frac{c_\rho}{\Phi_I}
\right)^{1/2} \left( A^{3/2} - 1 \right)\right] \, ,
\end{equation}
where we have used the initial value $A = 1$. Most moduli particles
decay when $\Phi \simeq \Phi_{I} \,\exp(-1)$, which occurs at
$A \simeq \tilde A$ with
\begin{equation} \label{decaystart}
\tilde{A} = \left[ \frac{3}{2} \left( \frac {\Phi_I} {c_\rho} \right)^{1/2} 
+1 \right]^{2/3} \, .
\end{equation}
Writing
\begin{equation} \label{gam}
H_I  = \gamma \Gamma_{\phi}\,,
\end{equation}
and using the fact that $H_I \propto \sqrt{\Phi_I}$, it is easy to see
that $\tilde A \gg 1$ requires $\gamma \gg 1$.

In this case, and with eqs.(\ref{initialc2}) imposed, the final
computed relic dark matter abundance will be essentially independent
of the initial $\Phi_I$, or, equivalently, of $\gamma$. Note that the
final scaled densities introduced in eqs.(\ref{eq:densities}) do
depend on $\Phi_I$. For example, again setting
$\widetilde{H} = \sqrt{\Phi_I}=$~constant a good analytical
approximation for the final value of $R$ can be derived \cite{Kane:2015qea}:
\begin{equation} \label{radbr}
R_F \equiv R(A_F) \simeq L \left( \frac{\Phi_I} {c_{\rho}} \right)^{1/3}\,
\Phi_I\,,
\end{equation}
where we have defined
\begin{equation}  \label{Beff} 
L \equiv  (1-B_{\rm eff})\, \Gamma\left(\frac{5}{3}\right)\, 
\left( \frac{3}{2}\right)^{2/3} \ \ {\rm with} \ 
B_{\rm eff} \equiv \frac{ B_{X^\prime} \left( {M_{X^\prime}}^2 + 3 {T_{\rm RH}}^2
\right)^{1/2}} {M_\phi}\, .
\end{equation} 
Here $A_F$ should be so large that the comoving abundances of
radiation and matter have become constant and $\phi$ decays have been
completed, i.e. $A_F \gg \tilde A$ with $\tilde A$ given by
eq.(\ref{decaystart}). On the other hand, $A_F$ should still be within
the radiation dominated epoch.

Evidently $R_F \propto \Phi_I^{4/3}$. This dependence cancels once we
normalize the final dark matter density to today's energy density
carried by photons, which is known very well from measurements of the
Cosmic Microwave Background (CMB). We thus compute the final
$X^\prime$ relic abundance from:\footnote{In ref.\cite{Kane:2015qea}
  the final DM relic density is expressed via today's radiation
  density. The latter is, strictly speaking, not known very well,
  since we do not know whether the lightest neutrino is still
  relativistic, and hence contributes to radiation, or
  not. Alternatively we could normalize to the entropy density, which
  was comoving constant for $A \geq A_F$; this closely mirrors the
  usual treatment of thermal WIMPs. Note also that we do know today's
  entropy density of neutrinos, which also remained comoving constant
  for $T \ll$ 1 MeV.}
\begin{eqnarray} \label{relicmod}
\Omega_{DM} h^2 &=& \frac{ \rho_{X^\prime}(T_{\rm now})} {\rho_\gamma(T_{\rm now})} 
\Omega_\gamma h^2
= \frac{\rho_{X^\prime}(T_F)} {2 \rho_R(T_F)} \frac {g_*(T_F)
h_*(T_{\rm now})} {h_*(T_F)} \frac{T_F} {T_{\rm now}} \Omega_{\gamma} h^2 
\nonumber \\ 
&=& M_{X^\prime} \frac {X^\prime(T_F)} {R(T_F)} \frac {A_F T_F g_*(T_F) 
h_*(T_{\rm now})} {2 T_{\rm now} T_{\rm RH} h_*(T_F)} \Omega_\gamma h^2 \, .
\end{eqnarray} 
Here $\Omega_{DM}$ is the dark matter mass density in units of the
critical density, $h$ is today's Hubble constant in units of
$100 \ {\rm km/(s\cdot MPc)}$, and $T_F = T(A_F)$ is in the radiation
dominated era, as mentioned above. In the second step we have written
$\rho_\gamma = 2 \rho_R/g_*$, and used the fact that the matter
density $\rho_{X^\prime}$ scales exactly like the entropy density
$s_R$ for $A \geq A_F$, i.e. after all $\phi$ decays and $X^\prime$
annihilations ceased. Note also that $h_*$ becomes constant after
electrons decoupled, i.e. for $T \ll m_e$. The present observational
values of the current temperature and density of (CMB) photons cosmic
microwave background (CMB), as collected by the Particle Data Group
\cite{Agashe:2014KDa}, are:
\begin{eqnarray} \label{eq:T_now}
\Omega_{\gamma} h^{2}&=& 2.473\times10^{-5}\,; \\ \nonumber
T_{\rm now}&=&2.7255 \ {\rm K}= 2.35\times 10^{-13}\ {\rm GeV}\,.
\end{eqnarray}
We use these values in our numerical calculations. Cosmological
observations also determine the total present density of non-baryonic
dark matter quite accurately \cite{Agashe:2014KDa}:
\begin{equation} \label{eq:Om_now}
\Omega_{DM} h^2 = 0.1186 \pm 0.002\,.
\end{equation}
This can be used to effectively reduce the dimension of the allowed
parameter space by one. However, in this paper we are more interested
in mapping out the predicted relic density as function of the relevant
free parameters. These include the reheat temperature, the branching
ratio for $\phi \rightarrow X^\prime$ decays, the $X^\prime$ annihilation
cross section as parameterized in eq.(\ref{eq:sigav}), and the masses
of the $\phi$ and $X^\prime$ particles. In Sec.~\ref{initdarkrad} we
will in addition allow non--vanishing initial values for the radiation
and $X^\prime$ densities.

\section{Dark Matter Relic Density for Initially Vanishing Radiation}
\label{dmprod}

In this Section we show numerical results for the predicted $X^\prime$
relic density for vanishing initial radiation and $X^\prime$ densities,
i.e.  for initial conditions given by eqs.(\ref{initialc1}) and
(\ref{initialc2}). All $X^\prime$ particles -- and all other particles
in today's universe -- then originate from $\phi$ decay, either
directly or via the radiation that originates from $\phi$ decay.

Before presenting numerical results, it is useful to briefly discuss
the different DM production mechanisms in non--thermal cosmology. Here
we again closely follow ref.\cite{Kane:2015qea}, where analytical
approximations based on eqs.(\ref{eq:boltzmann}) were developed. As
discussed in the previous Chapter, these equations ignore terms that
depend on the derivative of $g_*$ or $h_*$ with respect to
temperature. While our numerical treatment fully includes these
effects, the analytical approximations remain useful as a guide to the
(quite large) parameter space. We also remind the reader that, unlike
ref.\cite{Kane:2015qea}, we do not include a dark radiation component.

It should be clear that the usual thermal WIMP scenario can also be
reproduced in our framework, if $T_{\rm RH}$ is above the conventional
decoupling temperature $T_{\rm FO}$ defined in the radiation dominated
epoch. This requires rather large reheat temperatures, and hence very
large $\phi$ masses as shown in Fig.~1, or else quite small masses for
the dark matter particle $X^\prime$. In the latter case one would
typically need additional light mediators in order to achieve a
sufficiently large $X^\prime$ annihilation cross section. Of course,
here we are mostly interested in scenarios that differ from this
standard thermal WIMP scenario.

A first important observation is that for not too small $X^\prime$
annihilation cross section, the rhs of the third
eq.(\ref{eq:boltzmann}) essentially vanishes over an extended range of
$A$. In the absence of $\phi \rightarrow X^\prime$ decays this
corresponds to $X^\prime$ particles being in full thermal equilibrium,
but in the present context this ``quasi--static equilibrium'' (QSE)
can also be obtained through a balance between $X^\prime$ production
from $\phi$ decay and $X^\prime$ annihilation, with negligible
$X^\prime$ production from the thermal plasma.  The general QSE
solution is:
\begin{equation} \label{xpqse}
X^\prime_{\rm QSE}(A) = \left[\frac{A^3}{\langle \sigma v \rangle^{\prime}}
\left( \frac{c_{\rho}^{1/2} B_{X^\prime}} {c_{1}^{1/2} M_{\phi} M_{Pl}}\Phi \right) 
+ {X^\prime_{\rm EQ}}^2 \right]^{1/2}\,\,.
\end{equation} 
QSE will be maintained only if the reaction rate $n_{X^\prime} \langle
\sigma v \rangle^\prime$ is larger than the Hubble expansion rate. This
requires $X^\prime \geq X^\prime_{\rm crit}$, with
\begin{equation} \label{xpcrit}
X^\prime_{\rm crit} \equiv (n_{X^\prime})_{\rm crit} \frac{A^3}{T_{\rm RH}^3} =
\frac {HA^3} {\langle \sigma v \rangle ^\prime T_{\rm RH}^3}
= \frac{\widetilde{H}A^{3/2}}{c_{1}^{1/2} M_{Pl}T_{\rm RH}
 \langle \sigma v \rangle ^\prime}\,.
\end{equation} 
Clearly QSE can only be achieved if
$X_{\rm QSE}^\prime > X^\prime_{\rm crit}$. This leads to a {\em
  lower} bound $\langle \sigma v \rangle^\prime_c$ on the thermally
averaged $X^\prime$ annihilation cross section. This lower bound
depends on the temperature. If $T$ is much smaller than
$M_{X^\prime}$, the term $\propto \left(X^\prime_{\rm EQ} \right)^2$
in eq.(\ref{xpqse}) can be ignored. On the other hand, for
sufficiently high temperature $\langle \sigma v \rangle^\prime_c$
can be calculated from $X^\prime _{\rm QSE}=X^\prime _{\rm  EQ}$.
Explicit expressions for the critical cross section can be found in
\cite{Kane:2015qea}.

One can classify different regions of parameter space according to
whether the thermally averaged $X^\prime$ annihilation cross section is
above or below the critical one for $T \simeq T_{\rm RH}$; this is called
{\em efficient} and {\em inefficient} annihilation. 

We first consider the case of efficient $X^\prime$ annihilation. Let
$\hat T_{\rm FO}$ be the $X^\prime$ freeze--out temperature, computed
in a radiation dominated universe. If $\hat{T}_{\rm FO} > T_{\rm RH}$
then $X^\prime_{\rm EQ} \sim 0$ for $T \lsim T_{\rm RH}$. In this
``non--relativistic quasi static equilibrium'' case the relic abundance
can be approximated by
\begin{equation} \label{qse}
\Omega h^2[{\rm QSE_{nr}}] \propto \frac{ M_{X^\prime}} {g_*(T_{\rm RH})^{1/6} 
L^{3/4 } M_{\rm Pl} \langle\sigma v\rangle^\prime T_{\rm RH}}.
\end{equation}
In this case the final dark matter density depends on $\phi$
properties only via $T_{\rm RH}$. The dependence on the annihilation
cross section is as in the standard thermal WIMP scenario; however,
here the relic density is also proportional to the $X^\prime$ mass.

If $X^\prime$ annihilation is efficient at $T_{\rm RH}$ and 
$\hat{T}_{\rm FO} < T_{\rm RH}$ we are back in the standard scenario. The relic
density for non--relativistic and relativistic dark matter particles can
then be estimated as \cite{kt}:
\begin{equation} \label{foradnr}
\Omega h^2[{\rm FO^{rad}_{nr}}] \propto  \frac{1} {g_*(\hat{T}_{\rm FO})^{1/2}} 
\frac{\hat{x}^\prime_{\rm FO}} {M_{\rm Pl} \langle \sigma v\rangle^\prime }\,,\
{\rm with} \ 
\hat{x}^\prime_{\rm FO} = \frac{M_{X^\prime}} {\hat{T}_{\rm FO}}\,;
\end{equation}
\begin{equation} \label{foradr}
\Omega h^2[{\rm FO^{rad}_r}] \propto \frac{M_{X^\prime}} {g_*(\hat{T}_{\rm FO})}\, .
\end{equation}

If the annihilation of DM particles is inefficient at $T_{\rm RH}$,
the DM relic will be affected by $X^\prime$ production during the
early matter dominated era and the branching ratio for
$\phi \rightarrow X^\prime$ decay. Since most $\phi$ decays occur at
$T \sim T_{\rm RH}$ when $X^\prime$ annihilation is assumed to be
inefficient one can write \cite{Kane:2015qea}
\begin{equation} \label{eq:omsum}
\Omega_{X^\prime} h^2 = \Omega _{\rm ann} h^2 + \Omega _{\rm decay} h^2 \,.
\end{equation}
The contribution $\Omega _{\rm decay} h^2$ comes from $\phi$ decays and obeys
\begin{equation} \label{decayprod}
{\Omega _{\rm decay}} h^2 \propto L^{-3/4} B_{X^\prime} \frac{T_{\rm RH} M_{X^\prime}}
{M_\phi} \, .
\end{equation}

The second contribution to the rhs of eq.(\ref{eq:omsum}) stems from
the interactions of $X^\prime$ particles with the thermal plasma
during the matter dominated epoch. Recall that we are assuming these
interactions to be negligible at $T \sim T_{\rm RH}$. However, this
does not exclude the possibility that $X^\prime$ might have been in
equilibrium at higher temperatures, still in the matter dominated
epoch, and decoupled at temperature $T_{\rm FO}$ with
$T_{\rm max} > {T}_{\rm FO} > T_{\rm RH}$. This can happen only for
dark matter particles that were non--relativistic at decoupling
\cite{Kane:2015qea,Hamdan:2017psw}, i.e. $M_{X^\prime} > {T}_{\rm FO}$. This ``modified
non--relativistic freeze--out'' scenario leads to
\begin{equation} \label{fomod}
\Omega _{\rm ann} h^2[{\rm FO^{mod}_{nr}}] \propto \frac {g_*(T_{\rm RH})^{1/2}}
{L^{3/4} g_*(T_{\rm FO})} \frac {T_{\rm RH}^3 {x^\prime_{FO}}^4 }
{ {M_{X^\prime}}^3 M_{\rm Pl} \langle \sigma v\rangle^\prime}\,,\ {\rm with} \
x^\prime_{FO} = \frac{M_{X^\prime}}{T_{FO}}\,.
\end{equation}
Note that here the contribution to the relic density is again
inversely proportional to the annihilation cross section, as in the
case of standard thermal WIMPs. However, the dependence on
$T_{\rm RH}$ and $M_{X^\prime}$ is quite different (and stronger) than
in scenarios where $X^\prime$ annihilation is still efficient at
$T \sim T_{\rm RH}$, c.f. eq.(\ref{qse}).

On the other hand, for sufficiently small annihilation cross section
the $X^\prime$ density never reached equilibrium. As long as this
cross section is not zero, there will nevertheless be a contribution
to the dark matter relic density from $X^\prime$ pair production from
the thermal plasma.  This ``inverse annihilation'' contribution can be
significant both for relativistic ($M_{X^\prime} \ll T_{\rm RH}$) and
for non--relativistic ($M_{X^\prime} > T_{\rm RH}$) $X^\prime$ particles:
\begin{equation} \label{ianr}
\Omega _{\rm ann} h^2[{\rm IA_{nr}}] \propto \frac{g_*(T_{RH})^{3/2} T_{\rm RH}^7 M_{\rm Pl} 
\langle \sigma v\rangle^\prime} { g_*(T_{*})^3{M_{X^\prime}}^5 } \, ;
\end{equation}
\begin{equation} \label{iar}
\Omega _{\rm ann} h^2[{\rm IA_r}] \propto \frac{ T_{RH}M_{X^\prime} M_{\rm Pl} 
\langle \sigma v \rangle^\prime } {~g_*(T_{\rm RH})^{3/2}}\, .
\end{equation}
Note that this contribution is directly proportional to the
annihilation cross section (which equals the $X^\prime$ pair
production cross section); this is true also in standard cosmology if
the dark matter particles never attained equilibrium, e.g. because the
temperature was too low \cite{dik}. In the non--relativistic case the
dependence on $T_{\rm RH}$ and $M_{X^\prime}$ is very strong. The
production of $X^\prime$ particles that were non--relativistic at
$T_{\rm RH}$ peaks at $T_* \simeq 0.28 M_{X^\prime}$, when the dark
matter particles were in fact semi--relativistic. Note that eq. (\ref{ianr}) 
is valid only if $T_{\rm max}$ is larger than $T_*$; 
otherwise this contribution is exponentially suppressed. In contrast, the
production of relativistic $X^\prime$ particles peaks at
$T_* \simeq T_{\rm RH}/2$.

Altogether one can thus distinguish seven different $X^\prime$
production mechanisms: ${\rm FO^{rad}_{nr}}$, ${\rm FO^{rad}_r}$,
${\rm FO^{mod}_{nr}}$, ${\rm IA_r}$, ${\rm QSE_{nr}}$, ${\rm IA_{nr}}$ and
$\phi-$decay. They dominate in different regions of parameter
space. Of course, these regions are smoothly connected, i.e. one can
interpolate between these different regions.

\begin{figure}[htb]
\centerline{\includegraphics[width=0.8\textwidth]{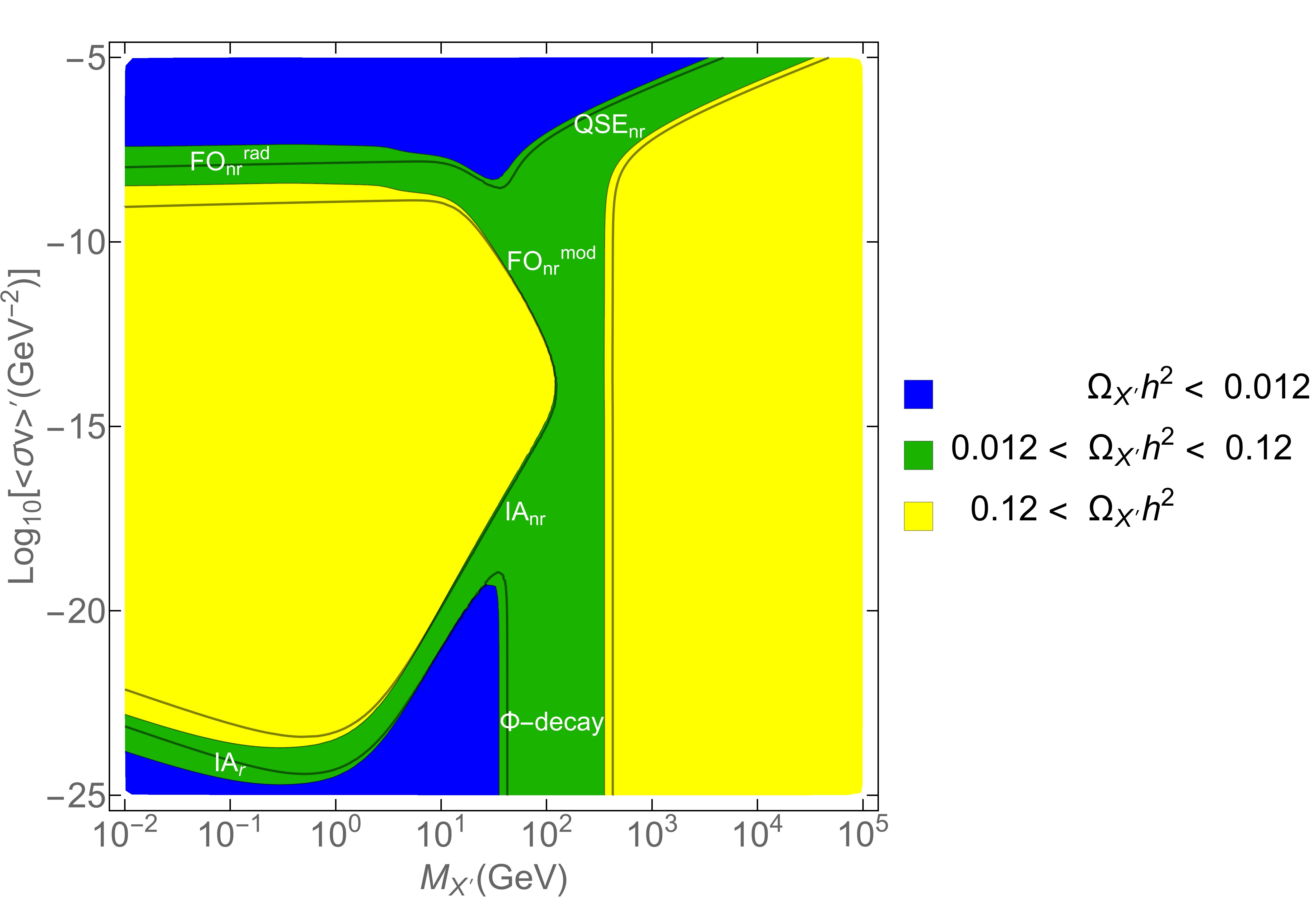}}
\caption{The dark matter relic density for $M_{\phi}= 5\times10^6$
  GeV, corresponding to reheating temperature $T_{\rm RH}=848.5$ MeV, $g_*(T_{RH})=73.46$, 
  and branching ratio $B(\phi \rightarrow X^\prime) = 10^{-5}$. The
  dark matter mass $M_{X^\prime}$ and the $S-$wave annihilation cross
  section $\langle \sigma v \rangle^\prime=a$ are given on the $x-$
  and $y-$axis, respectively. The colored regions represent different
  bins of the final dark matter relic density, computed including the
  full temperature dependence of $g_*$ and $h_*$, whereas the solid
  lines are contours of constant $\Omega_{X^\prime} h^2 = 0.12$ (deeper
  inside the yellow region) and $0.012$, respectively, under the
  approximation $g_* = h_* = g_*(T_{\rm RH})$.}
\label{dmcross1}
\end{figure}

Parameter regions where these different $X^\prime$ production
mechanisms are dominant are indicated in Fig.~\ref{dmcross1}. Here and
in the subsequent figures we use eq.(\ref{decaywidth}) with
$\alpha = 1$ to compute the total $\phi$ decay width, which in turn
determines the reheat temperature via eq.(\ref{eq:T_RH}). In this
figure we have assumed a rather heavy $\phi$ particle, and hence a
value of $T_{\rm RH}$ well above the temperature of the QCD
deconfinement transition. Moreover, we have assumed a constant
($S-$wave) $X^\prime$ annihilation cross section, and fixed
$B(\phi \rightarrow X^\prime) = 10^{-5}$. We do not consider
  values of $M_{X^\prime}$ below $10$ MeV, since for $M_{X^\prime} \ll T_{\rm RH}$
  the early $\phi-$matter dominated epoch becomes essentially
  irrelevant. On the other hand, we restrict ourselves to $M_{X^\prime}$
  values below a few percent of $M_\phi$ since otherwise the
  approximation of instantaneous thermalization of $\phi$ decay
  products might break down, as remarked above.

We see that for $M_{X^\prime} \leq T_{\rm RH}$ one recovers the
results of standard cosmology. In particular, in the top--left part of
Fig.~\ref{dmcross1} one recognizes the usual WIMP strip, where
$\Omega_{\rm DM} h^2$ comes out near the desired value if
$\langle \sigma v \rangle^\prime \sim 10^{-8}$ GeV$^{-2}$. Recall that
the freeze--out temperature in the radiation dominated epoch
$\hat T_{\rm FO}$ is about $20 M_{X^\prime}$; for
$\langle \sigma v \rangle^\prime \gsim 10^{-15}$ GeV$^{-2}$ large
deviations from standard cosmology therefore become evident only for
$M_{X^\prime} \geq 10$ GeV, which corresponds to
$\hat T_{\rm FO} \gsim T_{\rm RH}$. 

Still in the region of small $M_{X^\prime}$ another region with
roughly correct DM relic density can be seen for much smaller cross
sections. In this ``inverse annihilation'' region there is sufficient
$X^\prime$ pair production from the thermal plasma, but the $X^\prime$
density never reaches thermal equilibrium. The results of standard
cosmology are now only recovered for $M_{X^\prime} \lsim T_{\rm RH}$;
in this part of parameter space the ``inverse annihilation'' mechanism
can be considered to be an example of the freeze--in mechanism
\cite{Hall:2009bx}.

For larger $X^\prime$ the standard WIMP region merges into a region
where the correct relic density is obtained via thermal freeze--out in
the $\phi$ matter dominated epoch. Note that this requires
significantly {\em smaller} $X^\prime$ annihilation cross section than
in the WIMP region. The reason is that here the $X^\prime$ density
keeps getting diluted by the entropy produced by $\phi$ decays; recall
that the relic density is inversely proportional to the annihilation
cross section in both freeze--out regions, see eqs.(\ref{foradnr}) and
(\ref{fomod}). 

The latter of these equations also shows that in this region the cross
section required to obtain the desired relic density scales like
$M_{X^\prime}^{-3}$. As $M_{X^\prime}$ is increased the cross section
therefore rather quickly becomes too small for $X^\prime$ to achieve
full thermal equilibrium. Recall that the Hubble parameter in the
$\phi$ dominated epoch is (much) larger than in the radiation
dominated epoch at the same temperature, requiring a correspondingly
larger cross section to obtain equilibrium. Nevertheless in
Fig.~\ref{dmcross1} the region where the DM density for $X^\prime$
masses in the typical WIMP region (between 100 and 1000 GeV) comes out
roughly correctly extends to very small cross sections, the dominant
production mechanism being ``inverse annihilation'' or, for even
smaller $\langle \sigma v \rangle^\prime$, direct
$\phi \rightarrow X^\prime$ decays.

Finally, there is another region with roughly correct relic density in
Fig.~\ref{dmcross1}, where the $X^\prime$ annihilation cross section
is significantly {\em larger} than that required for thermal WIMPs in
standard cosmology. Here quasi--static equilibrium between $X^\prime$
production from $\phi$ decays and $X^\prime$ annihilation is achieved.
Eq.(\ref{qse}) shows that here the required cross section scales like
$M_{X^\prime}$. This region therefore merges with the ``modified
freeze--out'' region for $M_{X^\prime} \simeq 100$ GeV, but allows to
reproduce the correct relic density for very large $X^\prime$
annihilation cross sections if the $X^\prime$ mass is sufficiently
large.\footnote{The possibility to obtain the correct relic density in
moduli--dominated scenarios where the annihilation cross section is
too large for the normal thermal WIMP scenario was to our knowledge
first discussed in ref.\cite{moroi}.}

Note that in Fig.\ref{dmcross1} the DM relic density comes out roughly
correctly for $M_{X^\prime}$ of a few hundred GeV almost {\em
  independently} of the $X^\prime$ annihilation cross section, as long
as the latter does not exceed a few times $10^{-7}$ GeV$^{-2}$. This
remains true \cite{Kane:2015qea} also for lower $T_{\rm RH}$,
corresponding to lower $\alpha$ and/or lower $M_\phi$; however, the
$\phi$ decay region then directly merges into the QSE$_{\rm nr}$
region. Moreover, in Fig.~\ref{dmcross1} all regions with
approximately correct relic density are continuous. This is no longer
the case for lower $T_{\rm RH}$, where thermal effects are important
only for smaller $X^\prime$ masses; if the effective branching ratio
$B_{\rm eff}$ is kept fixed, there is then an extended region of
$X^\prime$ masses where the $X^\prime$ relic density is always too
low, independent of the $X^\prime$ annihilation cross section
\cite{Kane:2015qea}.

\begin{figure}[htb] 
\centerline{\includegraphics[width=0.8\textwidth]{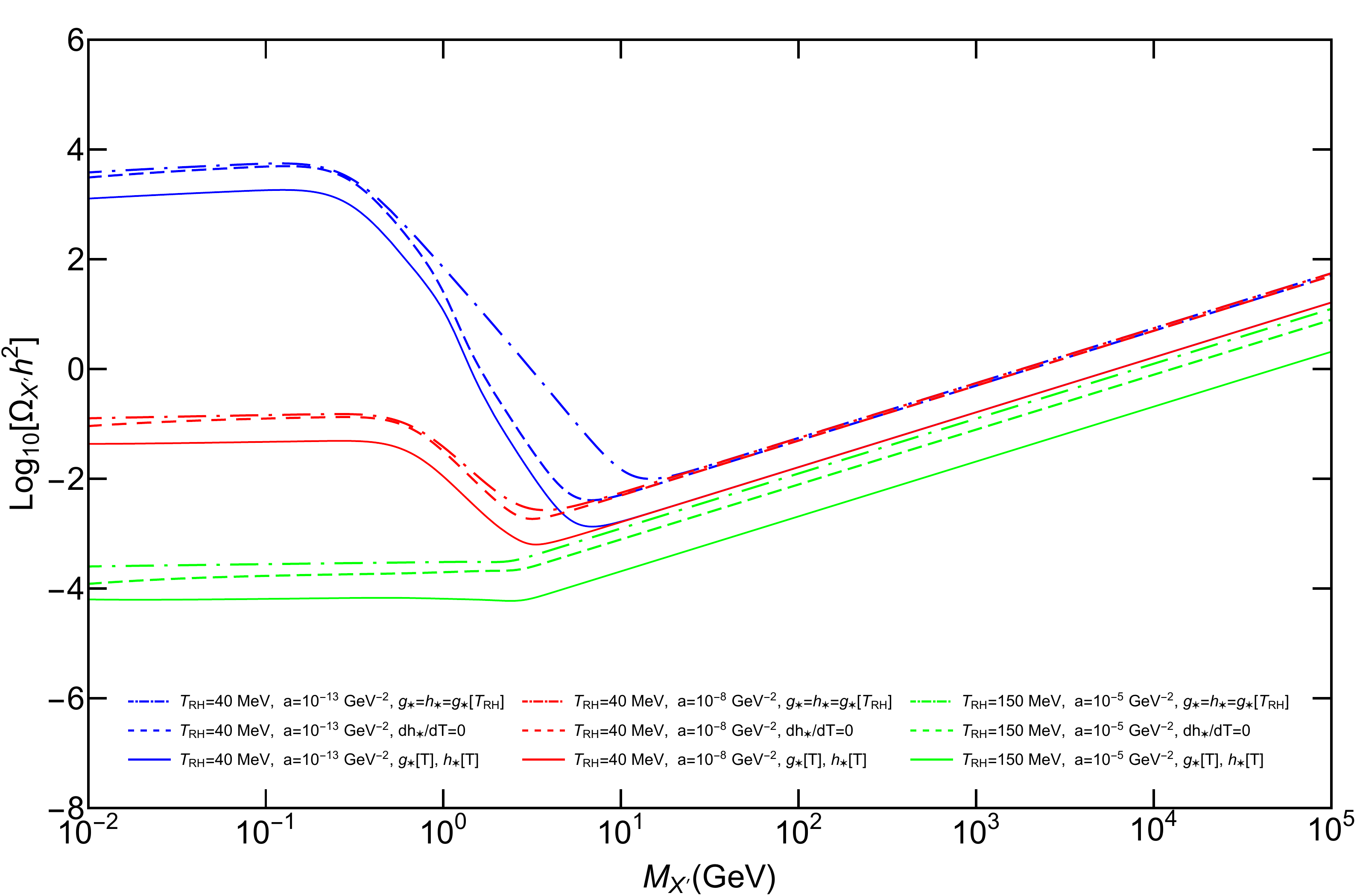}}
\caption{The predicted DM relic density as function of the DM mass
  $M_{X^\prime}$. Different colors refer to different choices of input
  parameters, as indicated in the frame. The dot--dashed curves have been
  obtained by setting $g_* = h_* = g_*(T_{\rm RH})$ everywhere. The
  other curves use a temperature dependent $g_*$ when calculating
  $\rho_R$, but the dashed curves have been obtained by setting
  $d h_* / dT = 0$.}
\label{different}
\end{figure}

These gross features are not affected by an accurate treatment of the
number of degrees of freedom in the thermal plasma. However, the solid
contours in Fig.~\ref{dmcross1} show that simply taking
$g_* = h_* = g_*(T_{\rm RH})$, which seems to have been the approach used in
ref.\cite{Kane:2015qea}, can lead to sizable errors of the final DM
relic density. This is further illustrated in Fig.~\ref{different},
where we show the predicted DM relic density as function of
$M_{X^\prime}$. The solid, dashed and dash--dotted curves have been
obtained by correctly treating the full temperature dependence of $g_*$
and $h_*$, by keeping $g_*$ and $h_*$ dependent on temperature but
setting $d\, h_* / d\, T = 0$ in eq.(\ref{temp2}), and by setting
$g_* = h_* = g_*(T_{\rm RH})$ everywhere, respectively.

We see that this last choice can over--predict the relic density by as
much as two orders of magnitude; see the blue (top) curves for
$M_{X^\prime} \simeq 3$ GeV. Here the relic density is determined by
freeze--out during the $\phi$ matter dominated epoch, with
$T_{\rm FO}$ not far from the QCD transition temperature where $g_*$
and $h_*$ vary quickly. In this example, $T_{\rm RH} = 40$ MeV is well
below the QCD transition, with $g_*(T_{\rm RH}) = 13.84$. Above the QCD
deconfinement transition the actual $g_*$ is much higher, which means
that the actual temperature is {\em lower} than that predicted in the
approximation $g_* =h_*= g_*(T_{\rm RH})$.

Moreover, setting the $d h_* / dT = 0$ over--predicts the relic
density by about a factor of three even for large $X^\prime$ masses,
where the relic density is set by direct $\phi \rightarrow X^\prime$
decays, which are independent of the thermal plasma. The reason is
that the final physical DM density is obtained by normalizing the
dimensionless comoving density $X^\prime$ to the radiation energy
density (or, equivalently, to the entropy density). Unlike in standard
cosmology, the comoving entropy density is {\em not} constant during
the epoch of $\phi$ matter domination; the actual temperature, or
entropy density, depends on the number of degrees of freedom in the
thermal plasma. Moreover, if one uses eq.(\ref{eq:s_R}) to compute the
entropy density, including the $T$ dependence of $h_*$ but setting
$d h_* / dT = 0$ or, equivalently, if one uses the second
eq.(\ref{eq:boltzmann}) to describe the evolution of the radiation
component, the entropy density $s_R$ will {\em not} be conserved in
the radiation--dominated epoch after $\phi$ decay.

\begin{figure}[htb] 
\centerline{\includegraphics[width=0.82\textwidth]{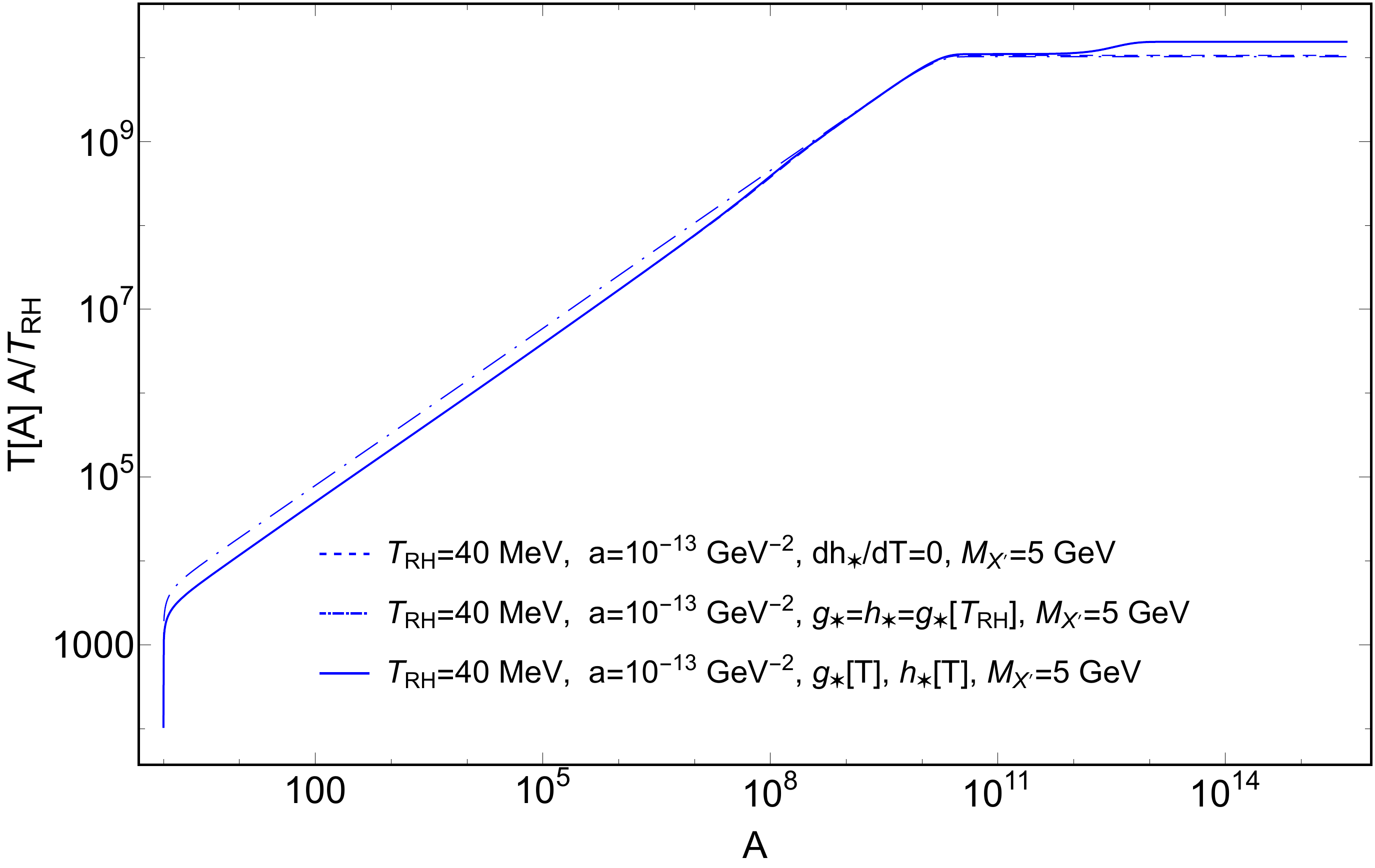}}
\caption{Evolution of the scaled temperature with A.}
\label{fig:temp}
\end{figure}

This is further illustrated by Fig.~\ref{fig:temp}, where we plot the
rescaled dimensionless temperature $\bar T = T A /T_{\rm RH}$ as a
function of $A$ for $T_{\rm RH} = 40$ MeV and
$H_I =10^{15} \Gamma_\phi$, which determine $\Phi_I$ via
$\Phi_I = 3 H_I^2 M_{\rm Pl}^2/(8 \pi T_{\rm RH}^4)$. Note that
$\bar T$ approaches a constant in the radiation dominated 
epoch if
$g_*$ and $h_*$ are constant. Since we assume initial temperature
$T_I = 0$, see eq.(\ref{initialc2}), the universe goes through the QCD
transition twice in this example: once early on, during the rapid
heating phase culminating at the maximal temperature estimated in
eq.(\ref{tmax}), and then again for much larger $A$, but (in this
example) still in the $\phi$ matter dominated epoch. Since
$d h_*(T) / dT \geq 0$ everywhere, the prefactor on the rhs of
eq.(\ref{temp2}) always tends to slow down the evolution of $T$, or
$\bar T$, with $A$. This implies a slower increase of $T$, and hence a
reduced $T_{\rm max}$, during reheating, but also a slower decline of
$T$ when the universe undergoes the QCD transition for a second
time. In particular, the simplified treatment with
$g_*(T) = g_*(T_{\rm RH})$ will considerably overestimate the
temperature, and hence thermal $X^\prime$ production, as long as
$T > T_{\rm QCD}$, as remarked above.

Note that in the $\phi$ matter dominated epoch the radiation content
of the Universe is basically determined by $\phi$ decays occurring in
the previous ${\cal O}(1)$ Hubble times; the radiation produced even
earlier is quickly redshifted and becomes irrelevant after a few
Hubble times. Hence $\rho_R(T)$, or $T$ itself, basically depends on
$g_*$ and $h_*$ only at temperatures $T^\prime \simeq T$. Therefore
the curves in Fig.~\ref{fig:temp} essentially coincide in the range of
temperatures where $g_*(T) \simeq g_*(T_{\rm RH})$.

Finally, the curves diverge again at very large $A$, well after all
$\phi$ particles have decayed. This is due to the decoupling of
$e^+e^-$ pairs, which increases the photon temperature by a factor
$1.4$ relative to a calculation where this effect is ignored. Of
course, in the case at hand one could have chosen to terminate the
numerical solution of the evolution equations at a value of $A_F$ such
that $T_F > m_e$ while still satisfying $T_F \ll T_{\rm RH}$. Still,
this feature shows that an accurate description of the evolution of
the universe in scenarios with a $\phi$ matter dominated epoch
requires a careful treatment of the temperature dependence of $g_*$
and $h_*$ over the entire range of temperatures.

The upshot of this discussion is that a simplified treatment that
ignores the temperature dependence of $g_*$ and $h_*$ will produce
reliable results only if the final temperature $T_F$ is 
chosen that
$g_*(T_F) \simeq g_*(T_{\rm RH})$, {\em and} if thermal $X^\prime$
production mechanisms are irrelevant at all temperatures $T$ where
$g_*(T) \neq g_*(T_{\rm RH})$. The former condition can only be
satisfied if $g_*$ remains essentially constant for an extended range
of temperatures around $T_{\rm RH}$, which in particular is not the
case if $T_{\rm RH}$ is near the QCD transition temperature. Since the
``(modified) freeze--out'' and the ``inverse annihilation''
contributions to the $X^\prime$ relic density depend on some range of
temperatures $T > T_{\rm RH}$ the question whether the second
condition is satisfied depends on several parameters
($T_{\rm RH}, M_{X^\prime}, \langle \sigma v \rangle^\prime,
B_{X^\prime}$) in a rather complicated manner.

\begin{figure}[htb] 
\centerline{\includegraphics[width=0.85\textwidth]{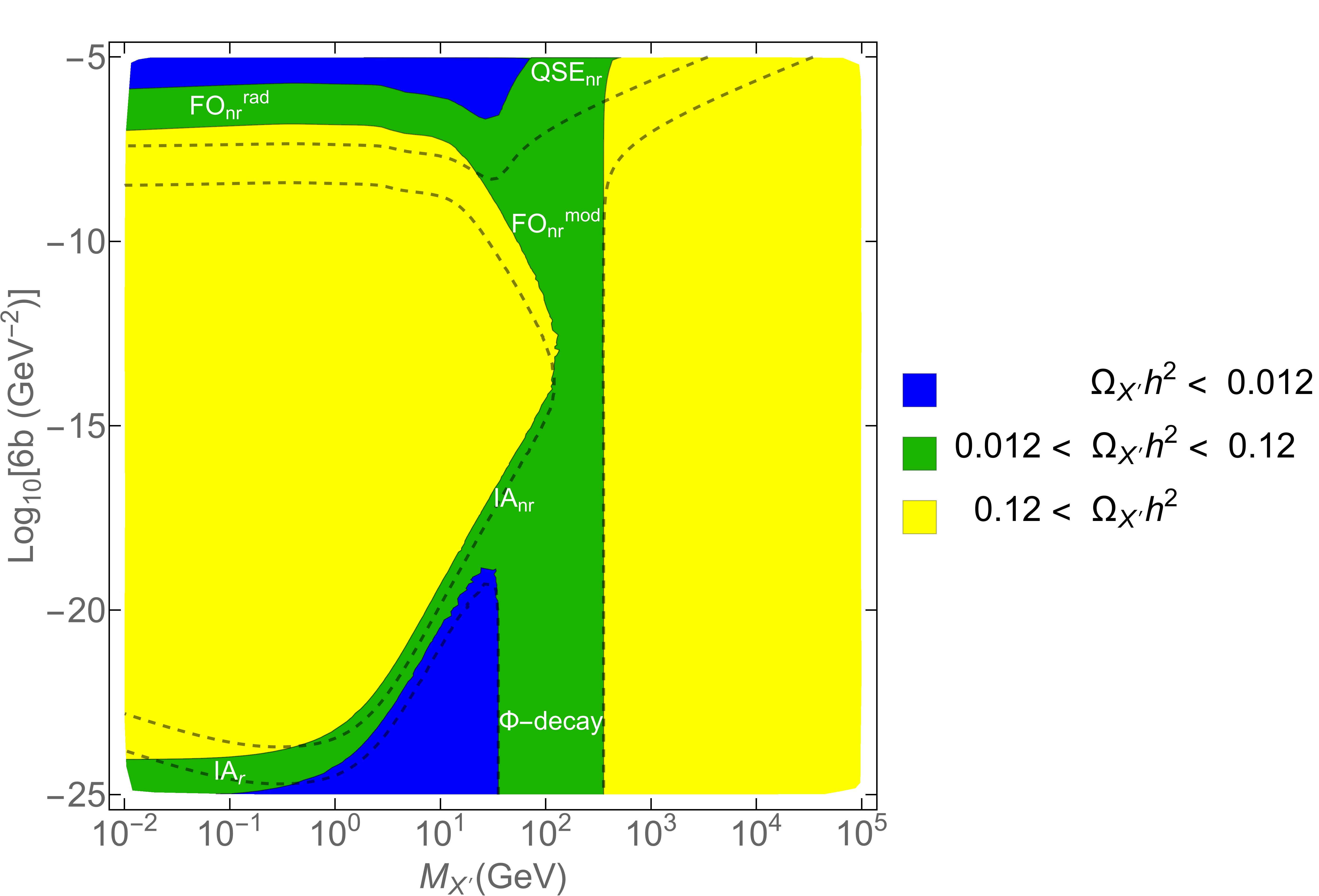}}
\caption{Contours of different values of the DM relic density with
  mass $M_\phi= 5\times10^6$ GeV, corresponding to $T_{\rm RH}=848.5$
  MeV. The dashed lines correspond to $\Omega_{X^\prime} h^2 = 0.12$ (for
  the lines deeper inside the yellow region) and $0.012$ assuming a
  constant cross section $\langle\sigma v \rangle^\prime=a$, whereas
  the colored regions have been obtained assuming a constant parameter
  $6b$ in $\langle\sigma v \rangle^\prime=6b T/M_{X^\prime}$.}
\label{swavep}
\end{figure}

So far we have assumed the thermally averaged $X^\prime$ cross section
to be a constant. This is a good approximation for non--relativistic
$X^\prime$ particles annihilating dominantly from $S-$wave initial
states. In Fig.~\ref{swavep} we compare this to results assuming
$\langle\sigma v \rangle^\prime= 6b T/M_{X^\prime}$ with constant
$b$. This reproduces the correct temperature dependence for
non--relativistic particles annihilating from a $P-$wave initial
state. Since $T/M_{X^\prime} \simeq 0.05$ for freeze--out in the
radiation--dominated epoch, $6b$ needs to be more than one order of
magnitude larger than $a$ in order to obtain the correct relic density
in the usual WIMP scenario. The difference between the allowed regions
is much less in the green strip to the right, where thermal effects
are either irrelevant ($\phi-$decay region) or peak at temperatures
not far from $M_{X^\prime}$ (inverse annihilation region); the one
exception occurs in the QSE region, where the relevant temperature
again satisfies $T \ll M_{X^\prime}$. The biggest change occurs in the
relativistic inverse annihilation region. In fact, using a constant
($T-$independent) annihilation cross section for $M_{X^\prime} \ll T$
is unphysical; if $X^\prime$ particles annihilate via the exchange of
mediators whose mass exceeds $T$, one instead expects
$\langle \sigma v \rangle^\prime \propto T^2$, i.e. an even stronger
$T-$dependence. The difference in slope between the green strip and the
region between the dashed curves at small $X^\prime$ masses and small
cross sections therefore indicates that the treatment used in
ref.\cite{Kane:2015qea} is not reliable here. However, since this
concerns the region of parameter space that is not affected by the
early $\phi-$dominated epoch, we will not pursue this issue any further.

\begin{figure}[htb]
\centering
\subfloat[]{
\includegraphics[width=55mm]{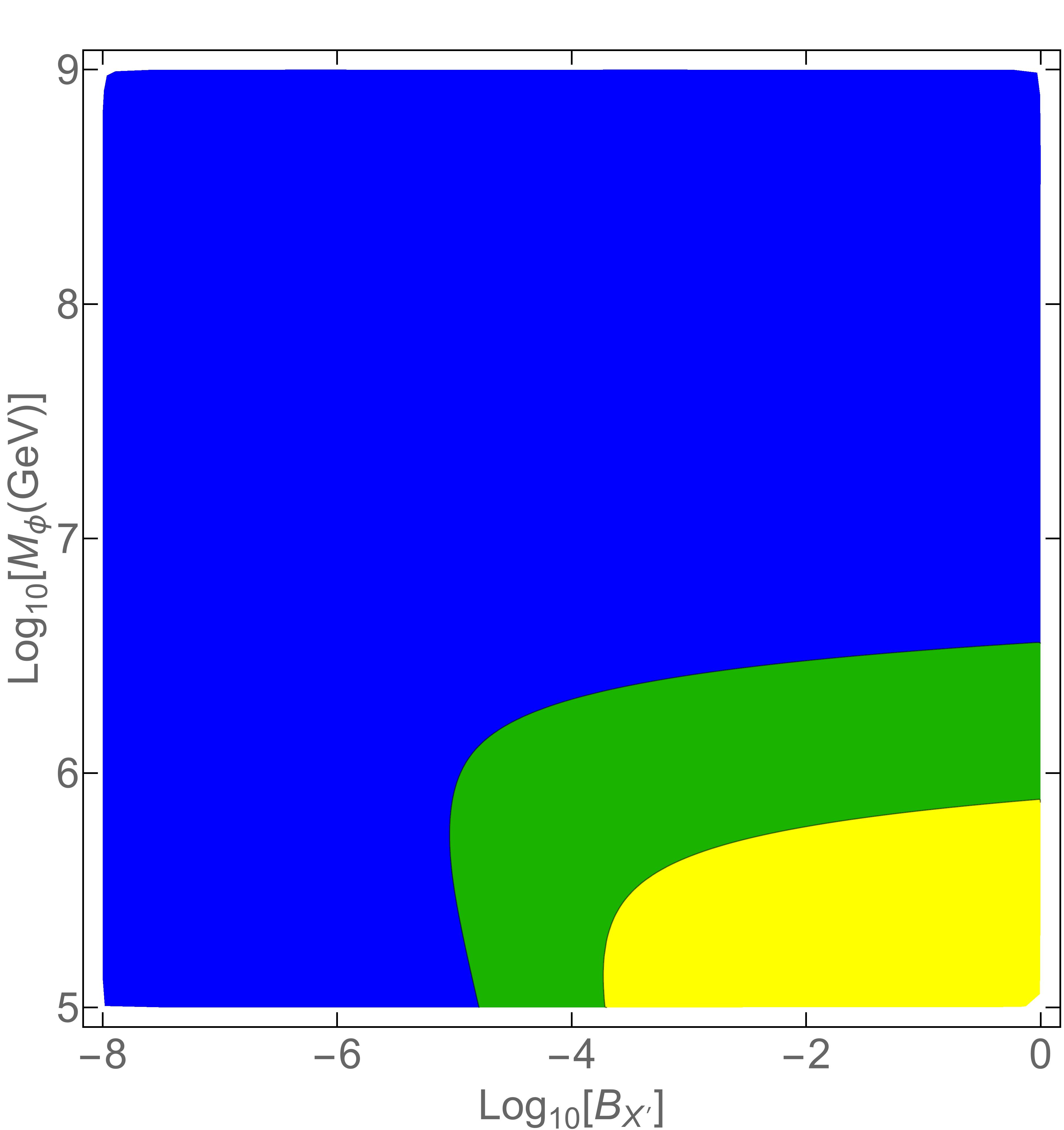}
\label{dmbranching1}
}
\subfloat[]{
\includegraphics[width=55mm]{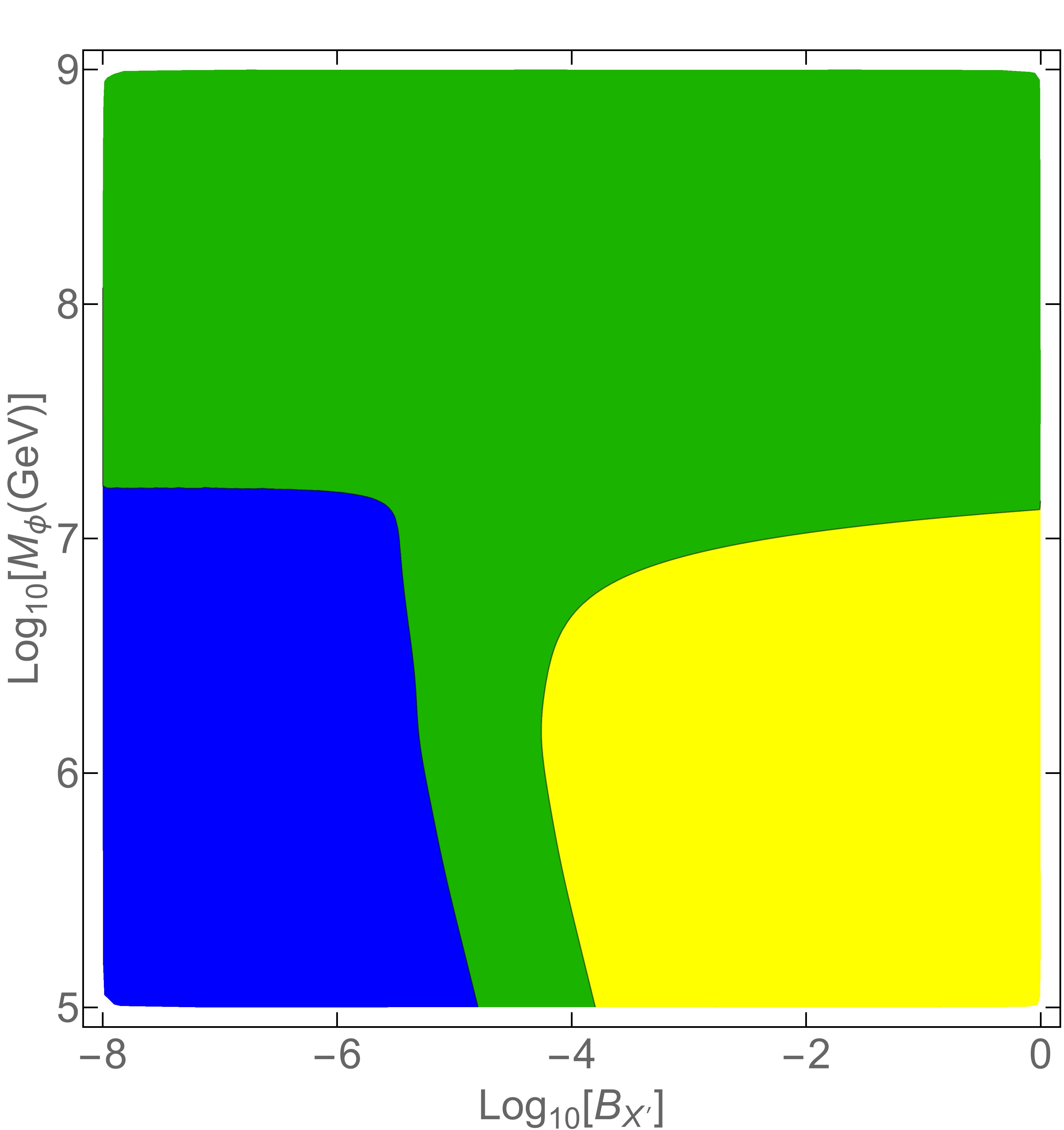}
\label{dmbranching2}
}
\hspace{0mm}
\subfloat[]{
\includegraphics[width=55mm]{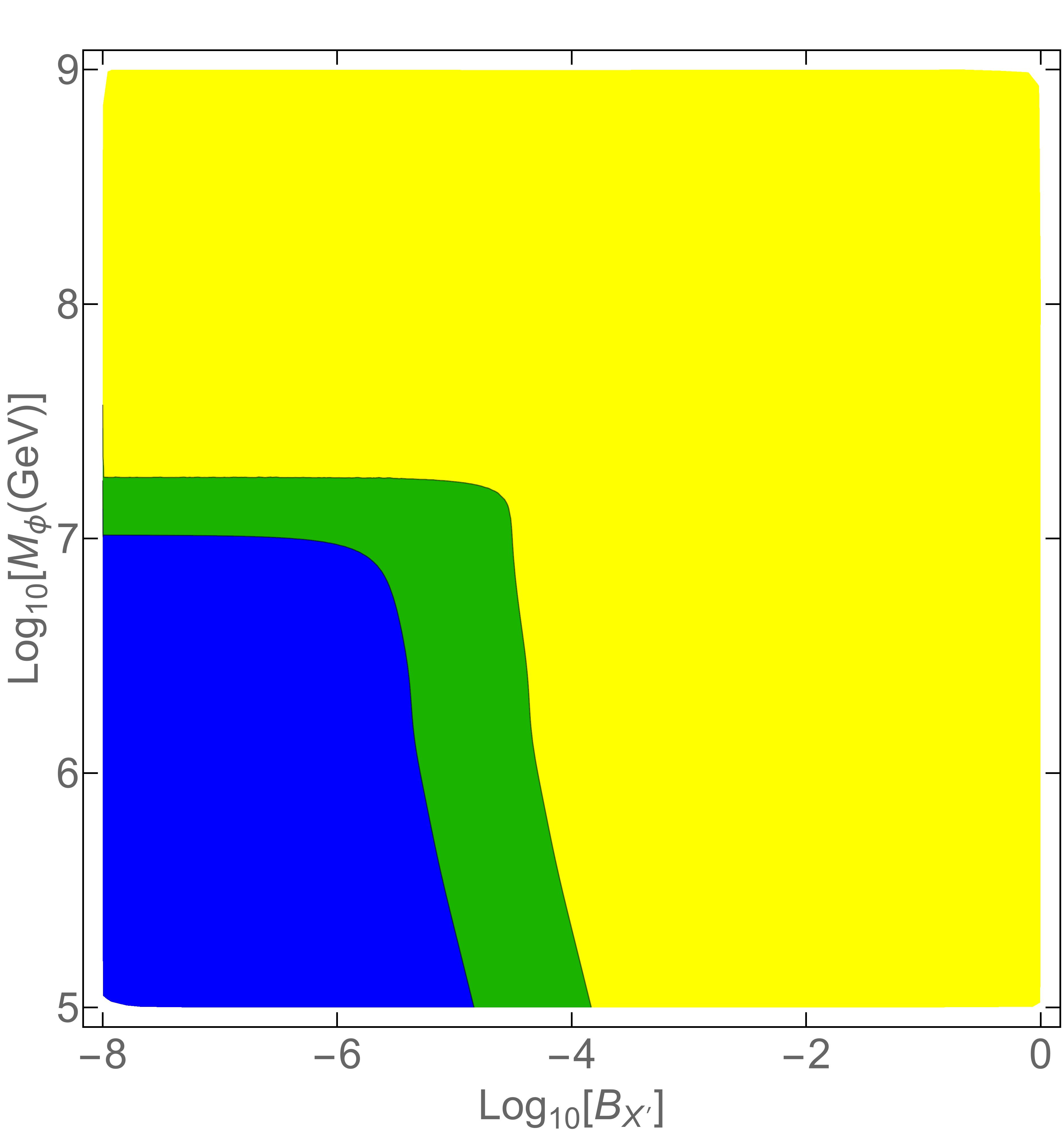}
\label{dmbranching3}
}
\subfloat[]{
\includegraphics[width=55mm]{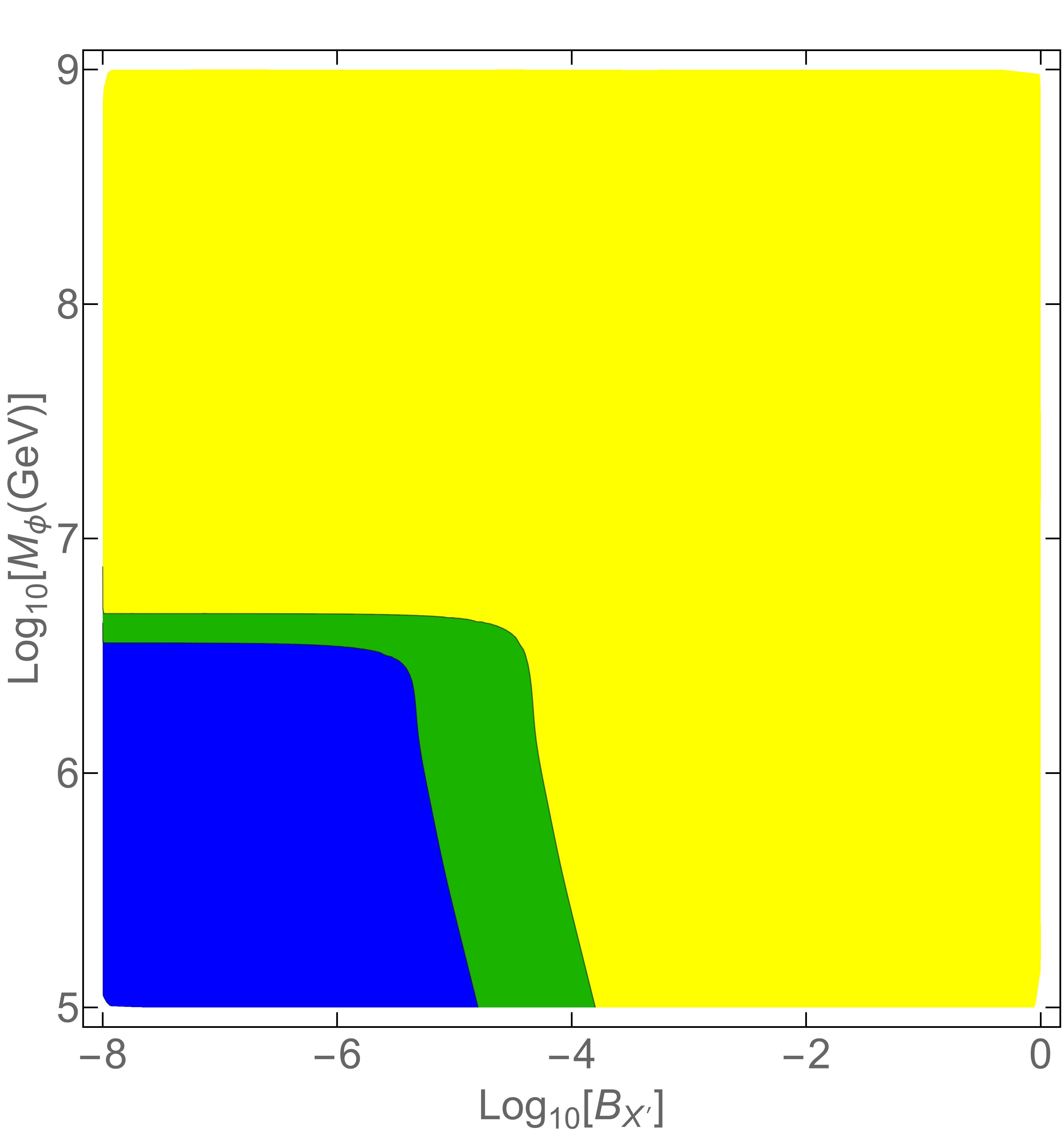}
\label{dmbranching4}
}
\hspace{0mm}
\subfloat[]{
  \includegraphics[width=55mm]{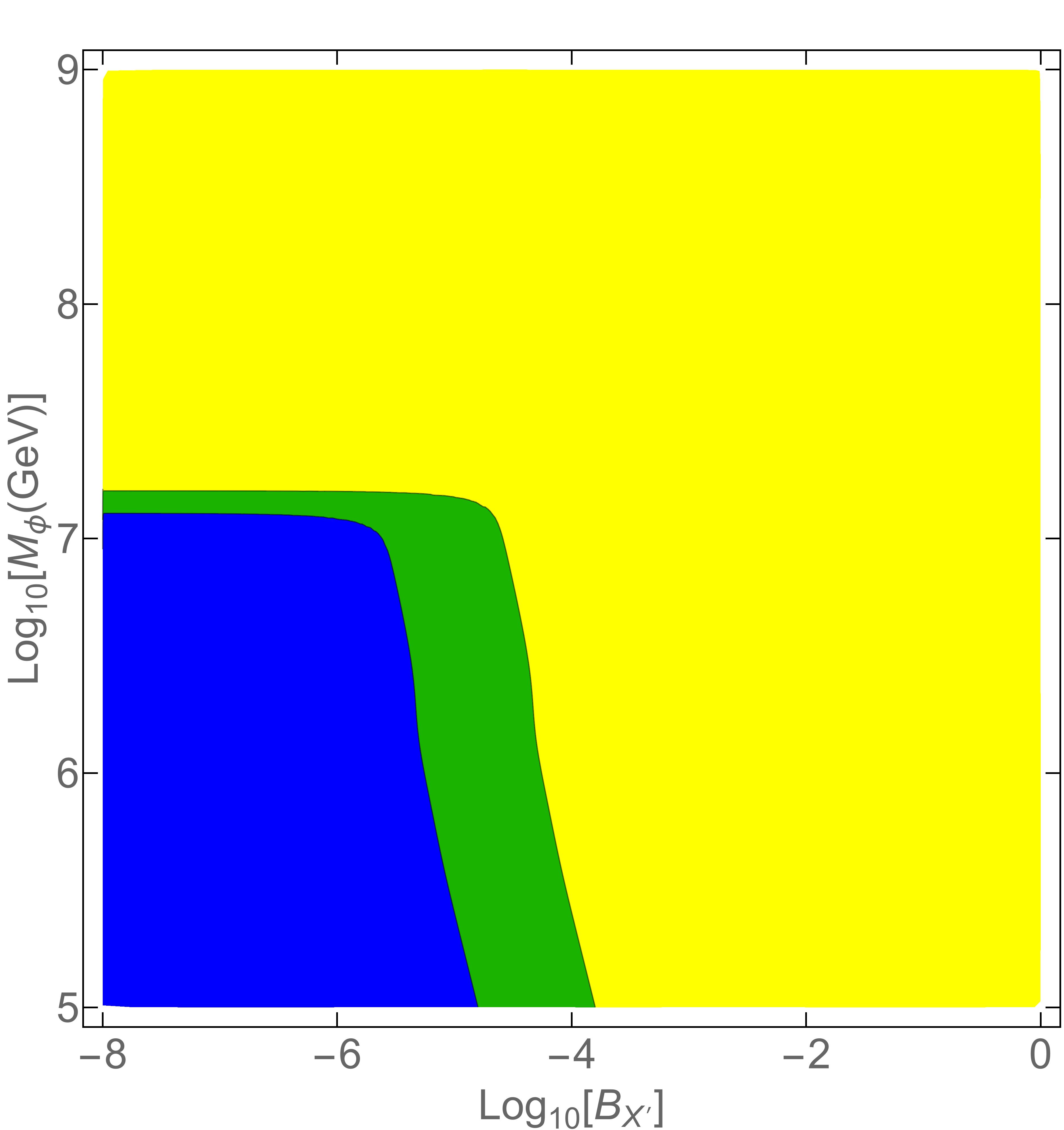}
\label{dmbranching5}
}
\subfloat[]{
\includegraphics[width=55mm]{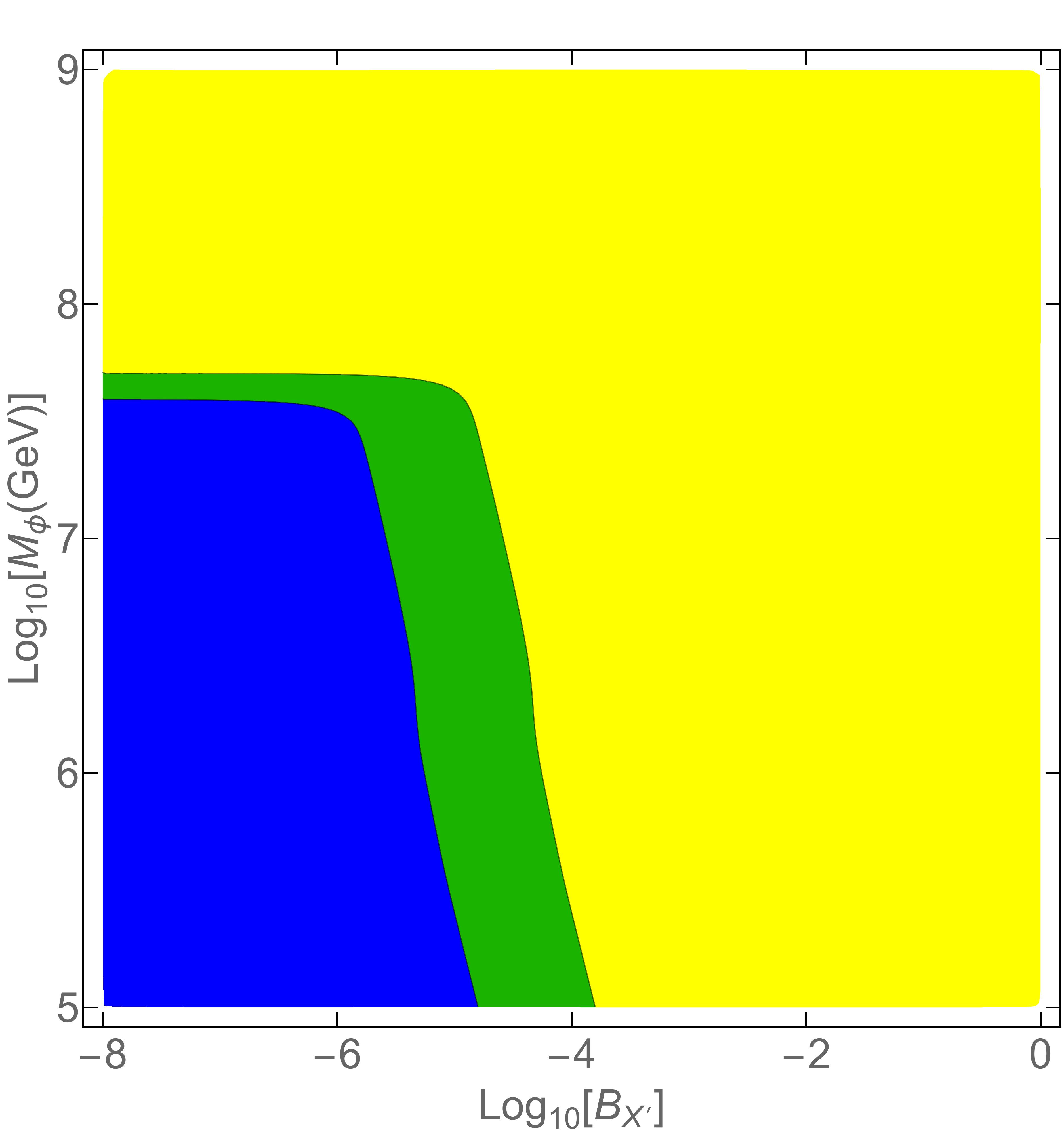}
\label{dmbranching6}
}
\caption{Contours of different values of DM relic density in the plane
  spanned by the modulus mass $M_\phi$ and the
  $\phi \rightarrow X^\prime$ decay branching ratio $B_{X^\prime}$. We
  have fixed the dark matter mass to $M_{X^\prime}=100$~GeV and
  different thermally averaged cross sections, taken to be independent
  of temperature. The thermally averaged cross section
  $\langle \sigma v\rangle^\prime$ in the figures is a)
  $10^{-6}$~GeV$^{-2}$, b) $10^{-8}$~GeV$^{-2}$, c)
  $10^{-9}$~GeV$^{-2}$, d) $10^{-14}$~GeV$^{-2}$, e)
  $10^{-20}$~GeV$^{-2}$, f) $10^{-25}$~GeV$^{-2}$, respectively. These
  results have been obtained using a careful treatment of the
  temperature dependence of $h_*$ and $g_*$. The colors are as in
  Fig.~\ref{dmcross1}.}
\label{dmbranching}
\end{figure}

In Figs.~\ref{dmbranching} we explore the dependence of the DM relic
density on the $\phi$ mass and the effective branching ratio for
$\phi \rightarrow X^\prime$ decays. In these figures the temperature
dependence 
of $g_*$ and $h_*$ has been treated carefully, but for
simplicity we have assumed $\langle \sigma v \rangle^\prime = a$ to be
independent of temperature; the six frames correspond to different
values of $a$, with fixed $M_{X^\prime} = 100$ GeV (a typical value
for a WIMP). Note that we have used eq.(\ref{decaywidth}) with
$\alpha = 1$ to compute the total $\phi$ decay width, which in
turn determines the reheat temperature via eq.(\ref{eq:T_RH}); hence
$T_{\rm RH}$ scales like $M_\phi^{3/2}$ in these figures.

In frame (a) we have chosen a rather large $X^\prime$ annihilation
cross section. Consequently the relic density is very low, unless
$M_\phi$ is rather small (so that $T_{\rm RH}$ is well below
$\hat T_{\rm FO}$) {\em and} $B_{X^\prime}$ is sizable. One is then
in the ${\rm QSE_{\rm nr}}$ region of parameter space, where the 
relic density scales like $T_{\rm RH}^{-1} \propto M_\phi^{-3/2}$, see
eq.(\ref{qse}). Note that the cross section required to achieve
quasi--static equilibrium scales like $1/B_{X^\prime}$.

Recall that in this region of parameter space the relic density is
proportional to the inverse of the $X^\prime$ annihilation cross
section.  Hence the region with too high relic density is
considerably larger in frame (b), which has ten times smaller
$\langle \sigma v \rangle^\prime$. In fact, now the relic density is
in the cosmologically interesting range even in standard cosmology,
which explains the large green region at large $M_\phi$, where
$T_{\rm RH} \geq \hat T_{\rm FO}$.

In the four remaining frames the $X^\prime$ annihilation cross section
is below that required for a thermal WIMP in standard cosmology. The
final DM density will then always be too large if
$B_{X^\prime} > 10^{-4}$; for these small cross sections, there is no
mechanism to sufficiently reduce a large $X^\prime$ density produced
directly from $\phi$ decays. Note that even if $\phi$ particles do not
directly couple to $X^\prime$ particles, $\phi$ decays into two SM
particles plus two $X^\prime$ particles (or an $X^\prime \bar X^\prime$ pair, if
$X^\prime$ is not self--conjugate) will in general still occur
\cite{allah2}. However, the resulting branching ratio is expected to
correlate with $\langle \sigma v \rangle^\prime$, so that a small cross
section also implies a small branching ratio for these four--body
modes, since both processes depend on the coupling of $X^\prime$ to SM
particles.  
\begin{eqnarray} \label{eq:R_F}
R_F \simeq R_I + R_F(\mu=0) \,,
\end{eqnarray}
\FloatBarrier

Even if $B_{X^\prime} < 10^{-4}$, the relic density will be too large
for $X^\prime$ particles with annihilation cross section below that of
standard thermal WIMPs if $M_\phi$ is too large. Recall that large
$M_\phi$ imply large $T_{\rm RH}$ and hence (too) large contribution
to the $X^\prime$ density either from inverse annihilation or, for yet
larger $M_\phi$, from the standard thermal freeze--out scenario.

The former dominates in the green regions at small $B_{X^\prime}$ in
the last three frames of Fig.~\ref{dmbranching}. For the chosen DM mass
$M_{X^\prime} = 100$ GeV, we see that $M_\phi \lsim 10^{7}$ GeV is
required, unless the $X^\prime$ annihilation cross section is many
orders of magnitude below that of thermal WIMPs. For non--relativistic
$X^\prime$ particles the annihilation cross section often scales like
$M_{X^\prime}^{-2}$. In this case we find numerically that the upper
bound on $M_\phi$ scales roughly like $M_{X^\prime}^{2/3}$. This
agrees with the estimate of eq.(\ref{ianr}) for the contribution to
the DM relic density from inverse annihilation during the $\phi$
matter dominated epoch. The same equation also shows that the upper
bound on $M_\phi$ will scale like $M_{X^\prime}^{10/21}$, or roughly
like $\sqrt{M_{X^\prime}}$, if we keep the annihilation cross section
independent of $M_{X^\prime}$; whereas for fixed $X^\prime$ mass, the
lower bound on $M_\phi$ will scale like
$(\langle \sigma v \rangle^\prime)^{-2/21}$, which explains why the
blue region in the last three frames of Fig.~\ref{dmbranching} only
grows rather slowly even though the annihilation cross section is
reduced by more than $10$ orders of magnitude.

\section{Dependence on Initial Conditions}
\label{initdarkrad}

In the previous Section we had assumed that the radiation and
$X^\prime$ densities initially vanish exactly. This is completely
realistic only if $\phi$ is a (weakly coupled) inflaton decaying
purely perturbatively into $X^\prime$ particles and/or radiation. In
contrast, in moduli cosmology one assumes that inflaton decay first
reheats the universe as usual. However, $\phi$ attains a large value
during inflation, so that eventually its density dominates the total
energy density. In this case the temperature will not be zero at any
time after inflaton decay. Of course, it stands to reason that if the
epoch of $\phi$ domination is sufficiently long, the initial
temperature will not matter, so imposing eqs.(\ref{initialc2}) will be
a good approximation. In this Section we investigate quantitatively
what impact a non--vanishing radiation content can have.

Even if at some sufficiently early time the universe is radiation
dominated, $\rho_R > \rho_\phi$, eventually these two densities will
become equal if $\phi$ particles are sufficiently long--lived, since
the ratio $\rho_\phi / \rho_R$ increases proportional to the scale
factor $a$. For a short time after this, the total radiation density
will still be dominated by the ``primordial'' component. In this
adiabatic regime (in the notation of ref.\cite{Co:2015pka}) the
temperature $T\propto a^{-1}$ because of entropy conservation. In the
subsequent ``non--adiabatic regime'', most radiation already comes
from $\phi$ decay and $T\propto a^{-3/8}$ as in eq.(\ref{tempa}). Note
that (after inflaton decay) the temperature of the universe never
increases in this scenario, as already pointed out in ref.\cite{kt}. 

It would be tempting to simply define our ``initial'' time, and ``initial''
scale factor, such that $\rho_{\phi,I} = \rho_{R,I}$. However, the case with
initially vanishing radiation density could then not be covered. Moreover,
at this initial time our dimensionless scale factor $A$ would usually not
be equal to $1$. We therefore prefer to define our initial conditions
such that $A=1$, and describe the initial radiation density through
the dimensionless parameter
\begin{eqnarray}
\mu = \frac{\rho_{R , I}}{\rho_{\phi,I}} \,.
\end{eqnarray}
The case covered in the previous Chapter obviously corresponds to
$\mu=0$, but very large (positive) values of $\mu$ are in principle
possible. We assume that the energy density of dark matter particles
is initially negligible compared to $\rho_\phi + \rho_R$. This should
be a good approximation even if the initial temperature
$T_I \geq M_{X^\prime}$ and $X^\prime$ particles were in full
equilibrium, simply because the total number of relativistic degrees
of freedom should be much larger than $g_{X^\prime}$. The Hubble
parameter is then given by
\begin{equation}
H^2_I = \frac{8 \pi}{3 M_{\rm Pl}^2}\left( \rho_{\phi ,I} + \rho_{R \,,I}\right)
= \frac{8 \pi \Phi_{I} T_{\rm RH}^4}{3 M_{\rm Pl}^2}  \left( 1 + \mu \right)\,.
\end{equation}
In our numerical examples we take $H_I$ (in units of $\Gamma_\phi$)
and $\mu$ as free parameters. The initial dimensionless comoving
densities of scalar and radiation can then be written as:
\begin{eqnarray} \label{initialmod}
\Phi_I &=& \frac{3 M_{\rm Pl}^2 H^2_I}{8 \pi T_{\rm RH}^4 \left( 1 + \mu \right)}
 \,, \nonumber \\
R_I &=& \mu \Phi_{I} \, .
\end{eqnarray}
The initial temperature can therefore be defined as
\begin{equation} \label{initialtemp}
T_I = T_{\rm RH} \left(\frac{30}{ \pi^2  g_*(T_I)} R_I \right)^{\frac{1}{4}} \,.
\end{equation}
In our numerical analyses we take $g_*(T_I) = 106.75$, which is the
number of degrees of freedom in the Standard Model if $T \gg m_t$ (top quark mass).

After the initial time, but before most $\phi$ particles have decayed,
the dimensionless Hubble parameter $\widetilde{H}$ can be estimated as
\begin{equation} \label{eq:H_A}
\widetilde{H} \simeq \Phi_{I}^{\frac{1}{2}} \left(1 + \frac{\mu}{A} 
\right)^{\frac{1}{2}} \,,
\end{equation}
where we again have neglected the contribution from $X^\prime$ particles.
Evidently the second term on the rhs of eq.(\ref{eq:H_A}) becomes
negligible once $A \gg \mu$; in this epoch the universe is again matter
dominated. Recall, however, that $\phi$ particles do eventually
decay at $A \simeq \tilde A$, see eq.(\ref{decaystart}). For $\mu \gsim 1$,
the $\phi$ matter dominated epoch therefore occurs for
\begin{eqnarray} \label{emd1}
\mu \ll A \lesssim \left(\frac{3}{2}\frac{\gamma}{(1+\mu)^{1/2}}+1 
\right)^{2/3} \, ,
\end{eqnarray}
where $\gamma = H_I / \Gamma_\phi$, see eq.(\ref{gam}). If $\mu > 1$,
an extended period of $\phi$ matter domination therefore requires
$\mu^2 \ll \gamma$.

On the other hand, $\gamma$ cannot be arbitrarily large in the
post--inflationary universe. We certainly need $H < M_{\rm Pl}$ in
order to treat gravity classically, see e.g. \cite{Linde:1983gd}.  In
inflationary cosmology the smallness of the density perturbations, and
the upper bound on primordial gravitational waves, requires
$H \lesssim 10^{-5} M_{\rm Pl}$ during inflation \cite{kt}, and hence
also afterwards. We therefore adopt the bound
$H_I < 10^{-5} M_{\rm Pl}$, which implies
\begin{equation} \label{gambound}
\gamma < \frac{10^{-5}}{\alpha} \left( \frac{M_{\rm Pl}}{M_{\phi}} \right)^3 \,.
\end{equation}
The rhs of (\ref{gambound}) is therefore also an upper bound on $\mu^2$
if the universe is to undergo a $\phi$ matter dominated epoch.

Since $X^\prime$ production or annihilation has little effect on the
thermal plasma, the final radiation density is simply given by
where $R_F(\mu=0)$ has been given in eq.(\ref{radbr}). If
$\mu^2 \ll \gamma$, the first term on the rhs of eq.(\ref{eq:R_F}) is
negligible, i.e. if the universe underwent an extended period of
$\phi$ matter domination, the final radiation density will come mostly
from $\phi$ decays.

However, the initial conditions may affect the final DM relic density
even if there is an epoch of $\phi$ matter domination. So far we have
only specified the initial radiation density in terms of $\mu$. In
complete generality the initial $X^\prime$ density is another free
parameter. However, we wish to avoid proliferation of parameters, and
therefore write the initial (co--moving, dimensionless) $X^\prime$
density as
\begin{equation} \label{eq:X_I}
X^\prime_I = \left( \frac{1} {T_{\rm RH}} \right)^3 \frac{g_{X^\prime} T_I 
{M_{X^\prime}}^2} {2 \pi^2} K_2 \left( \frac{M_{X^\prime}} {T_I} \right)\,.
\end{equation}
This is based on the Maxwell--Boltzmann distribution, but it is still
a reasonably good approximation for bosons and fermions from
relativistic to non--relativistic limits, as long as $X^\prime$ is
(approximately) in full thermal equilibrium with the hot plasma. This in
turn should be true if $T_I \gsim M_{X^\prime}$ unless the $X^\prime$ annihilation
cross section is very small: equilibrium should be reached if
\begin{equation} \label{equil}
g_{X^\prime} \langle \sigma v \rangle^\prime \gsim \left( \frac 
{ g_*(T_I) (1+\mu) } {30 \mu} \right)^{3/4} \pi^{7/2} \frac {1}
{\sqrt{ \alpha \gamma M_\phi^3 M_{\rm Pl} } }\,.
\end{equation}
Here we have again written $H_I = \gamma \Gamma_\phi$ and used 
eq.(\ref{decaywidth}) for $\Gamma_\phi$. The condition (\ref{equil}) can
only be violated if either $\mu$ or $\langle \sigma v \rangle^\prime$ is
very small. In the former case eq.(\ref{eq:X_I}) in any case predicts a
very small initial $X^\prime$ density, so it doesn't matter that this
small number may not be correct. In the latter case interactions of
$X^\prime$ with the thermal plasma will certainly remain negligible
at later times, so we can write the final $X^\prime$ as sum of the initial
value [which may not be given by eq.(\ref{eq:X_I}) then] and a possible
contribution from direct $\phi \rightarrow X^\prime$ decays:
\begin{equation}
X^\prime _F \simeq X^\prime_I + X^\prime_{F(,Br)} ,\ \ {\rm with}\ X^\prime_{F(,Br)} 
= \frac{B_{X^\prime}T_{\rm RH} \Phi_{I}} {M_{\phi}} \, .
\end{equation}

\begin{figure}
\centering
\subfloat[]{
\includegraphics[width=58mm]{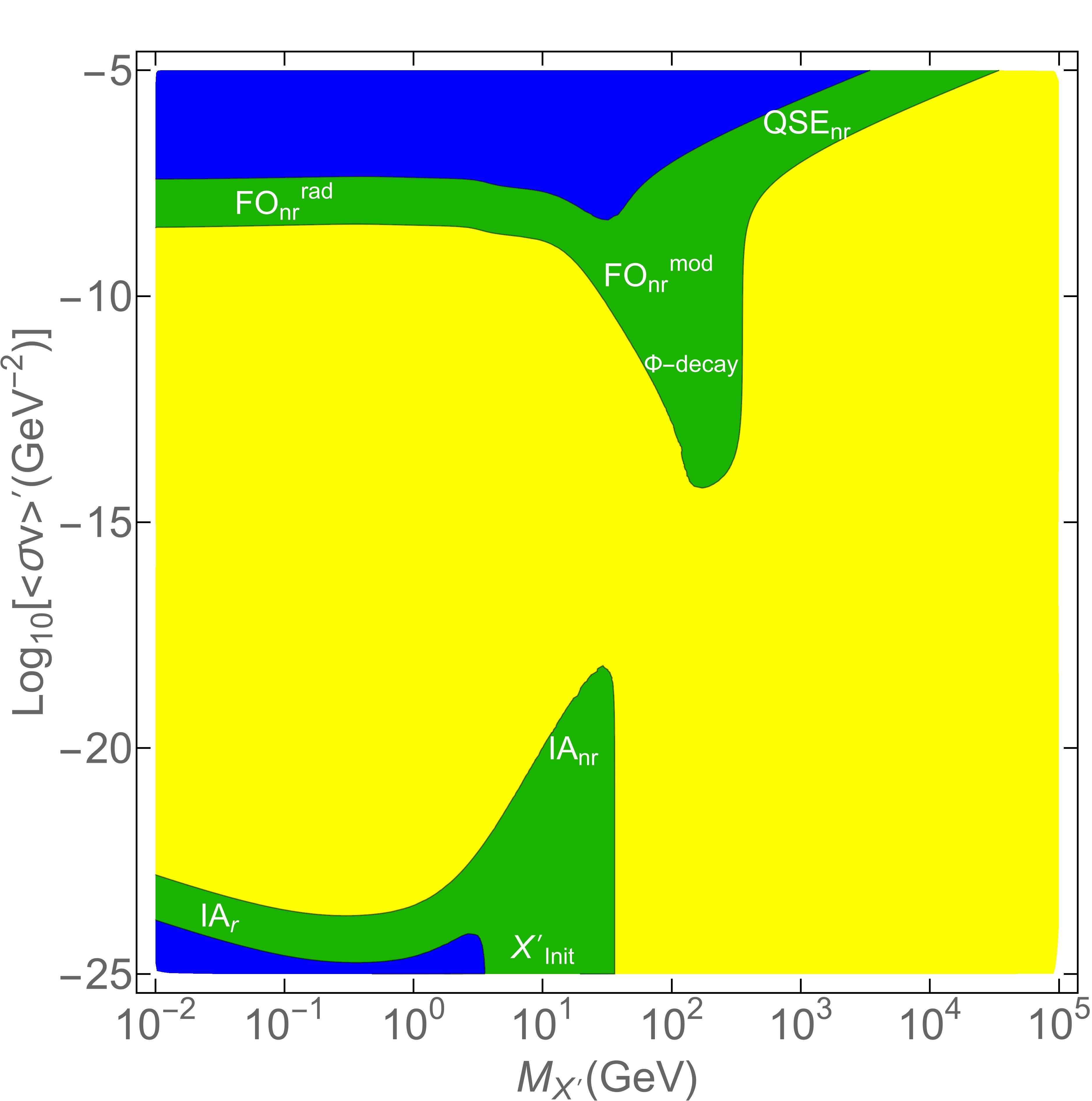}
}
\subfloat[]{
  \includegraphics[width=58mm]{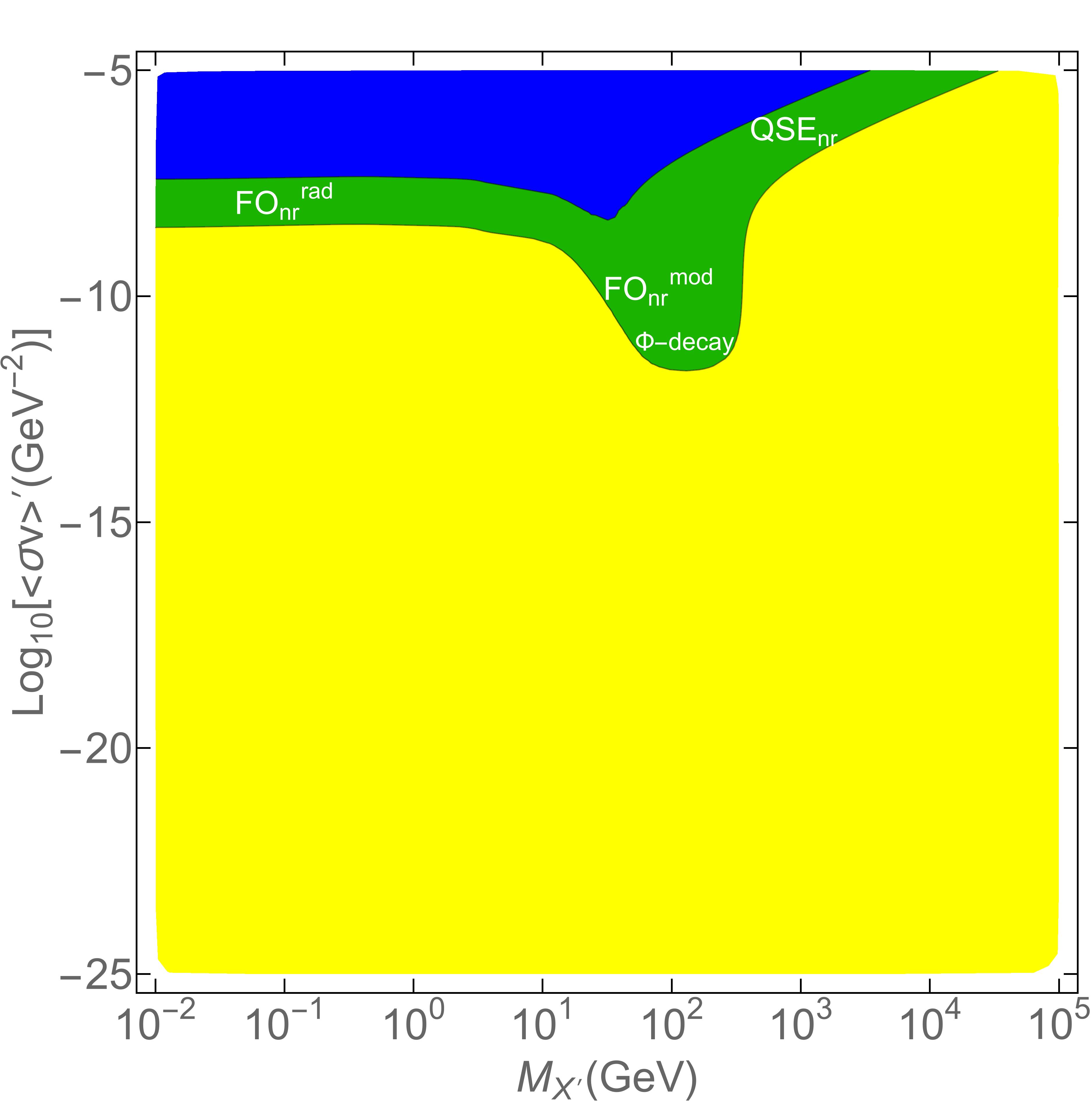}
}
\hspace{0mm}
\subfloat[]{
  \includegraphics[width=58mm]{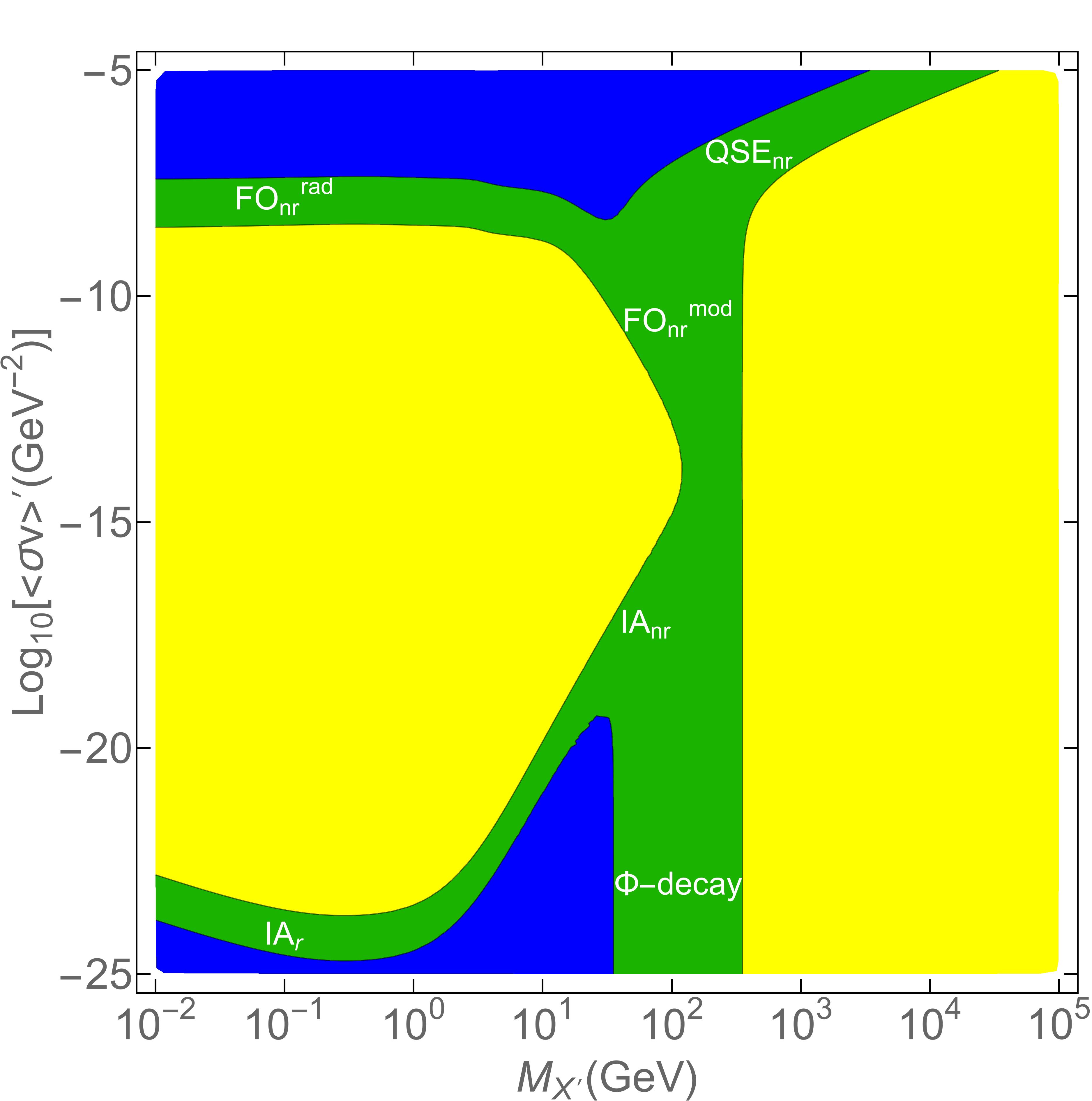}
}
\subfloat[]{
  \includegraphics[width=58mm]{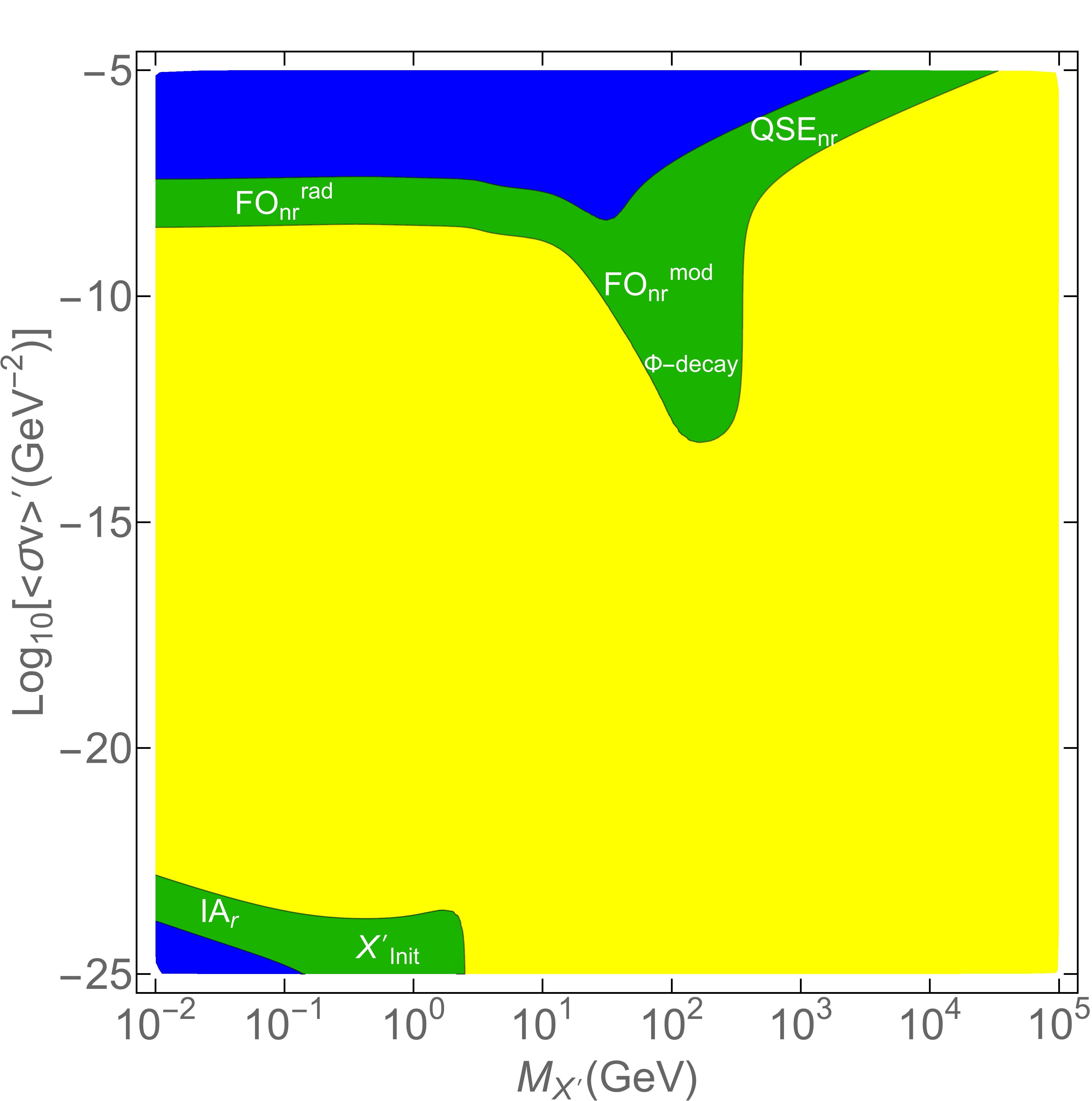}
}
\hspace{0mm}
\subfloat[]{
  \includegraphics[width=58mm]{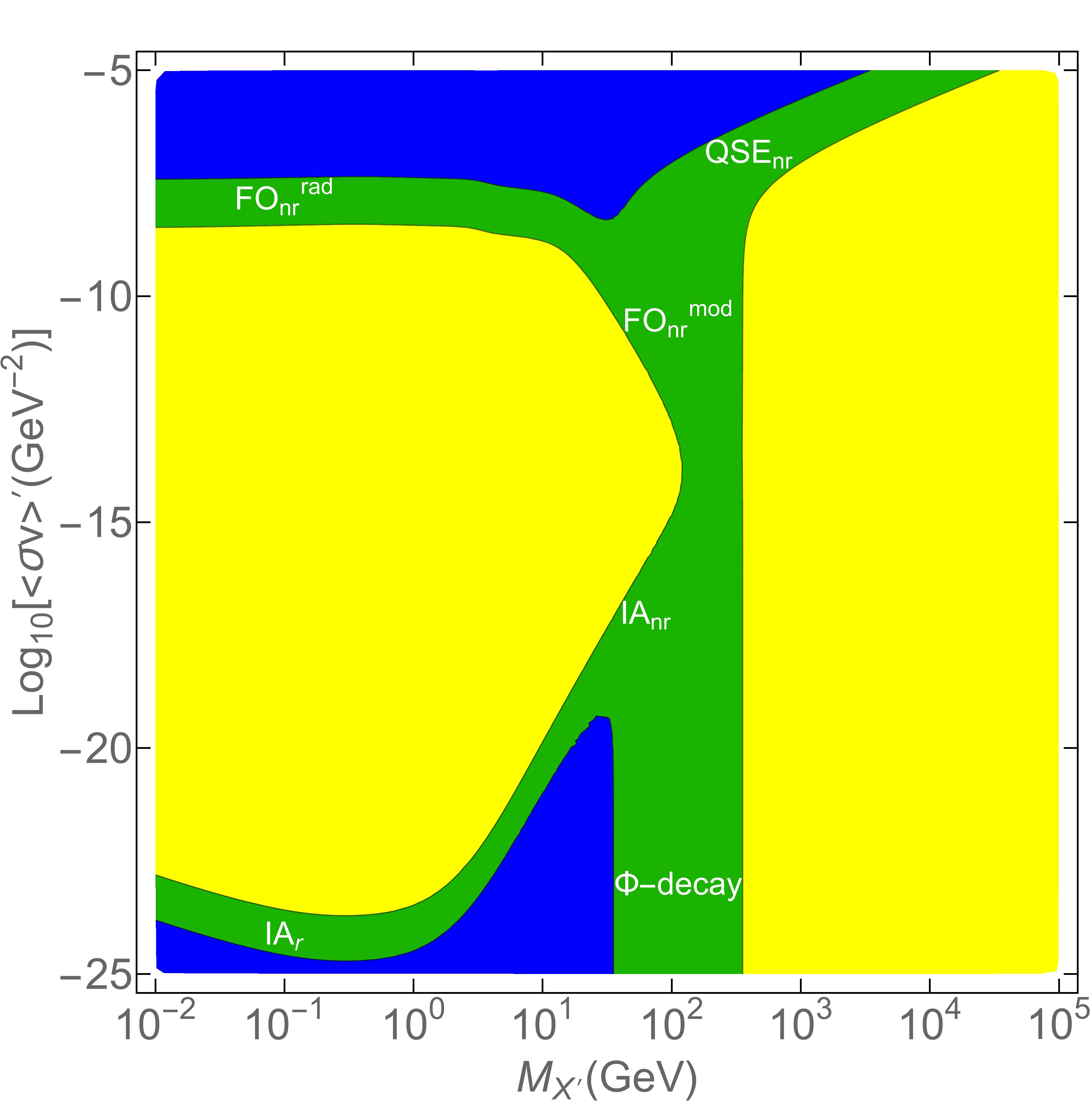}
}
\subfloat[]{
  \includegraphics[width=58mm]{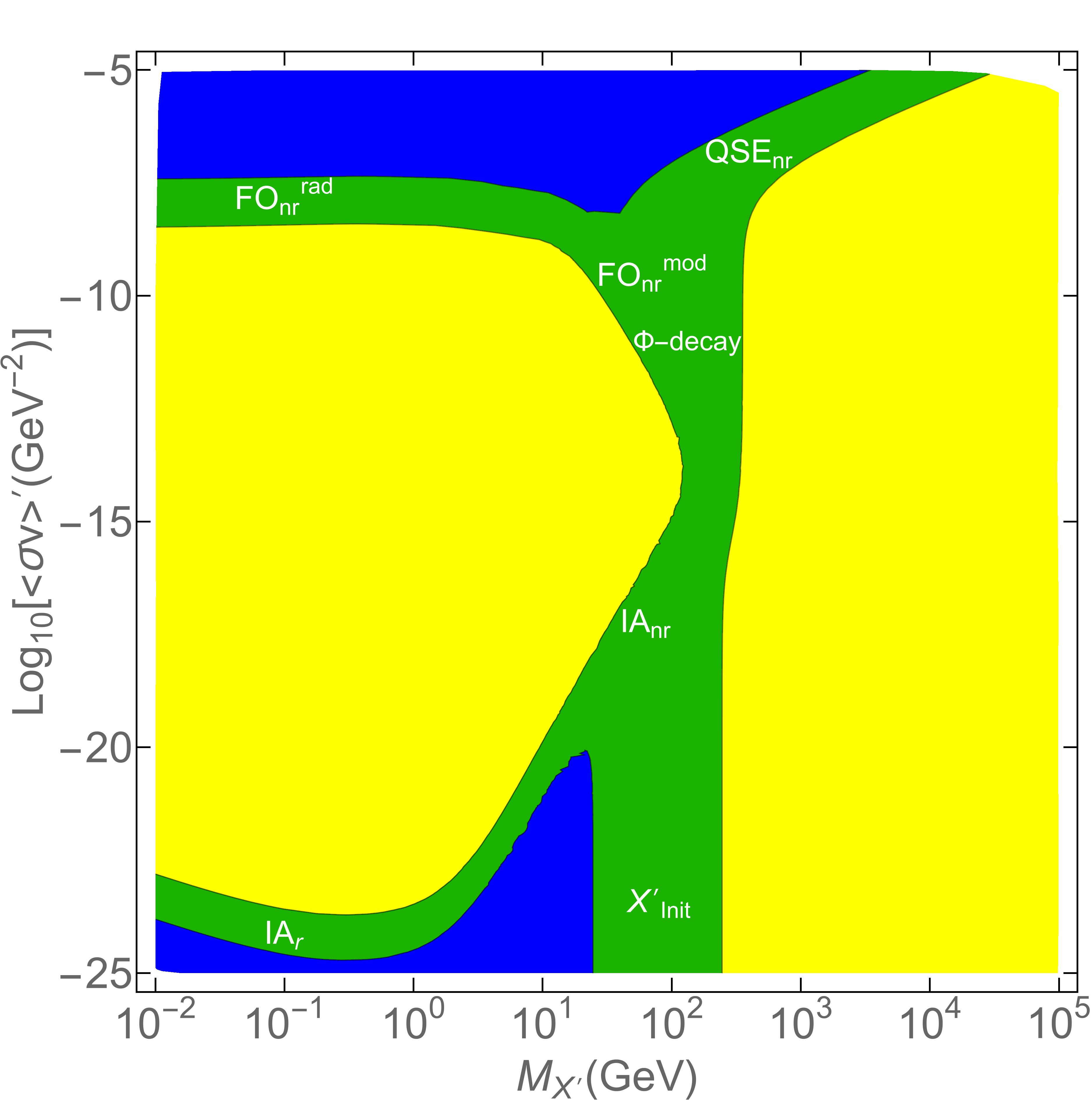}
}
\caption{Contours of constant relic density for different initial
  conditions. The meaning of the differently colored regions is as in
  Fig.~2; note that we only show results where the evolution of the
  number of degrees of freedom with temperature has been treated
  carefully. We have taken $M_\phi = 5\times10^6$~GeV and
  $\alpha = 1$, leading to $T_{\rm RH} = 848.5$~MeV, and 
  $B_{X^\prime} = 10^{-5}$. The six frames are for different
  combinations of $\gamma$ and $\mu$: $(\gamma, \, \mu) = $
  a) $(10^{10},\, 10^{-5})$, b) $(10^{10}, \,1)$, c) $(10^{15}, \, 10^{-5})$, d)
  $(10^{15}, \, 1)$, e) $(10^{20},\, 10^{-5})$, f) $(10^{20},\, 1)$.}
\label{initials}
\end{figure}

We are now ready to present some numerical results. In
Fig.~\ref{initials} we show the final DM relic density for the same
$\phi$ mass and $B_{X^\prime}$ as in Fig.~\ref{dmcross1}. We chose two
different values of the initial radiation (and $X^\prime$) density,
parameterized by $\mu$: $\mu = 10^{-5}$ (left column) and $\mu = 1$
(right column). Moreover, we chose three different values for the
initial Hubble parameter, parameterized by $\gamma$:
$\gamma = 10^{10}$ (first row), $\gamma = 10^{15}$ (second row), and
$\gamma = 10^{20}$ (third row). Since $\mu \leq 1$, in all examples
the universe is dominated by $\phi$ matter for all $A$ between $1$ and
$\tilde A$ defined in eq.(\ref{decaystart}). 
Note that condition (\ref{emd1}) is satisfied in all these cases.

We see that the initial conditions do not affect the final DM density
if the $X^\prime$ annihilation cross section is sufficiently large.
We saw in the previous Chapter that $X^\prime$ particles then achieve
full thermal equilibrium with the hot plasma during the epoch of
$\phi$ matter domination; for sufficiently high $T_{\rm RH}$,
$X^\prime$ will drop out of equilibrium only after all $\phi$
particles have decayed. Adding a non--vanishing initial radiation
component increases the temperature relative to the case $\mu = 0$,
making it easier for $X^\prime$ to attain thermal equilibrium. Hence
any scenario that leads to $X^\prime$ freeze--out for $\mu=0$ will have
$X^\prime$ in thermal equilibrium also for some time during $\phi$
domination. This period of thermal equilibrium will wipe out any
dependence of the final $X^\prime$ density on the initial conditions.
 For the parameters
of Fig.~\ref{initials} this is true for $\langle \sigma v \rangle^\prime
\gsim 10^{-12}$ GeV$^{-2}$.

\FloatBarrier
\newpage

Hence the initial conditions can affect the final DM density only if
for $\mu=0$ the latter is determined by the ``inverse annihilation''
or ``$\phi$ decay'' mechanisms discussed in the previous 
chapter, see
eqs.(\ref{decayprod}), (\ref{ianr}) and (\ref{iar}). The results
obtained for $\mu = 0$ will then only be approximately correct if the
initial value $X^\prime_I$ is much less than the final value of
$X^\prime$ produced during the epoch of $\phi$ matter domination.
From these equations and the initial condition (\ref{eq:X_I}) we find
that the initial contribution is negligible if:
\begin{eqnarray}  \label{ineq}
\frac{\mu^{3/4} (1+\mu)^{1/4}} {\gamma^{1/2}} &\ll& \kappa_{\phi-{\rm decay}} 
 B_{X^\prime} \left( \frac{ \alpha M_\phi }{ M_{\rm Pl} } \right)^{1/2} \,;
\nonumber \\
\frac{\mu ^{3/4}( 1 +\mu )^{1/4}}{\gamma^{1/2}} &\ll& {\kappa_{IA_{nr}}} 
\frac{\alpha^{1/2} M_\phi^{3/2} M_{\rm Pl}^{1/2} T_{\rm RH}^6 
\langle\sigma v\rangle^\prime} {M_{X^\prime}^6} \,;
\nonumber \\
\frac{\mu ^{3/4}(1+\mu)^{1/4}} {\gamma^{1/2}} &\ll& {\kappa_{IA_{r}}} 
\alpha^{1/2} M_\phi^{3/2} M_{\rm Pl}^{1/2} \langle\sigma v\rangle^\prime \,.
\end{eqnarray}
The first of these inequalities applies if $X^\prime$ production in
the epoch of $\phi$ domination is from direct
$\phi \rightarrow X^\prime$ decays, while the second and third
inequality apply if the main $X^\prime$ production mechanism for
$\mu = 0$ is inverse annihilation, with $X^\prime$ being
non--relativistic and relativistic, respectively. In these inequalities
we have only displayed the dependence on the free parameters; numerical
coefficients are collected in the $\kappa$'s, with $\kappa_{\phi-{\rm decay}} 
\simeq 35$, $\kappa_{IA_{nr}} \simeq 10^{-2}$. Altogether the
initial contribution to the $X^\prime$ density will be negligible if
the lhs is (much) less than the largest of the three right--hand sides.

When deriving these inequalities we have assumed that the initial
temperature is larger than $M_{X^\prime}$, so that
$X_I^\prime \propto T_I^3$ is not exponentially suppressed. Moreover,
we have assumed that the condition (\ref{emd1}) is satisfied, so that
the universe underwent an extended period of $\phi$ matter
domination. Finally, we have used eq.(\ref{decaywidth}) to compute
$\Gamma_\phi$; this is needed, since we express the initial Hubble
parameter, and hence $\Phi_I$, in terms of $\gamma$ defined in
eq.(\ref{gam}).

The first inequality (\ref{ineq}) is relevant for
$M_{X^\prime} \gsim 10$ GeV and
$\langle \sigma v \rangle^\prime \lsim 10^{-18}$ GeV$^{-2}$. For the
parameters used in Figs.~\ref{initials} the rhs amounts to about
$2\times10^{-10}$. The values of the lhs in the six frames are of order
$2 \times10^{-9}$ in (a), $10^{-5}$ in (b), $6 \times 10^{-12}$ in
(c), $3 \times 10^{-8}$ in (d), $2 \times 10^{-14}$ in (e), and
$10^{-10}$ in (f).  Correspondingly in the lower--right parts of the
plane shown in Fig.~\ref{initials} the initial contribution
$X^\prime_I$ dominates in (a), completely dominates in (b) and (d), is
subdominant but not completely negligible in (f) and can be neglected
in (c) and (e). The regions with approximately correct final DM
density which is dominated by $X^\prime_I$ are labeled as
$X^\prime_{\rm Init}$ in Figs.~\ref{initials}.

The situation is a bit more complicated in the part of parameter space
where $X^\prime$ production is dominated by inverse annihilation if
$\mu = 0$, since the rhs of the second and third inequalities
(\ref{ineq}) explicitly depend on the annihilation cross section and
-- for the second inequality -- the mass of the DM particle. Let us
focus on the region near the center of the plots, with
$M_{X^\prime} \simeq 50$ GeV and
$\langle \sigma v \rangle^\prime \simeq 10^{-17}$ GeV$^{-2}$, where
the inverse annihilation process produces approximately the correct
relic density for $\mu = 0$, with the DM particles being
non--relativistic already at production. The rhs of the second
inequality (\ref{ineq}) is of order $10^{-10}$ here. Correspondingly
in this central part of the parameter plane $X^\prime_I$ dominates in
frames (a), (b) and (d), is negligible in (c) and (e), and contributes
about equally in (f).

We thus see that for small DM annihilation cross section, one may need
$\gamma > 10^{20}$ in order to be independent of the initial
conditions, even if we require the initial radiation density to be not
larger than the initial $\phi$ mass density. In contrast, for
$\mu = 0$ the final DM relic density is independent of $\gamma$ once
$\gamma \gsim 10^7$. Note that for $\mu = 1$,
$T_I \sim \sqrt{\gamma} T_{\rm RH}$. Since BBN constraints imply
$T_{\rm RH} \geq 4$ MeV and $T_I$ should be smaller than the reheat
temperature after inflation, the latter would have to be at least
$10^8$ GeV if $\gamma \sim 10^{20}$. For the parameters of
Fig.~\ref{initials}, $\gamma > 10^{20}$ with $\mu = 1$ implies
$T_I \gsim 10^{10}$ GeV. Note, however, that possible problems from a
high post--inflationary reheat temperature, e.g. overproduction of
gravitinos, are alleviated by the huge amount of entropy produced
during the very long epoch of $\phi$ matter domination and
out--of--equilibrium decay of $\phi$ particles.

\section{Summary and Conclusions}
\label{conclusion}

This paper treats the production of Dark Matter particles in
cosmological scenarios with an early matter dominated epoch. This
occurs quite naturally in inflationary cosmology if the theory
contains a scalar particle $\phi$ with mass smaller than the Hubble
scale during inflation and with greatly suppressed couplings to SM
particles, and hence long lifetime. We improved on previous analyses
of this non--thermal DM production scenario by carefully treating the
temperature dependence of the number of relativistic degrees of
freedom ($g_*$ and $h_*$, defined via the energy density and entropy
density of radiation, respectively), and by investigating the effect
of a non--vanishing initial radiation and DM density.

We found that a careful treatment of the temperature dependence of
$h_*$ is very important over large regions of parameter space. This is
in sharp contrast to the more commonly considered scenario of WIMP
freeze--out in standard cosmology, where the $T-$dependence of $h_*$
only matters if $h_*$ varies rapidly around the freeze--out
temperature; and even then the effects do not exceed the $10\%$
level. In contrast, in the presence of an early matter--dominated
period approximating $h_*$ by a constant can lead to predictions that
are off by a large factor. One reason is that one always normalizes
the DM density to the radiation density, or equivalently to the
entropy density. In standard cosmology the comoving entropy density is
basically constant after the end of inflation. This is not the case in
the scenarios considered here, where (nearly) all of the entropy
density is produced from $\phi$ decays. An incorrect treatment of the
entropy density therefore immediately leads to a wrong prediction of
the final DM density. Moreover, some production mechanisms -- in
particular, the production of DM particles from the thermal plasma,
called ``inverse annihilation'' in ref.\cite{Kane:2015qea} -- are
quite sensitive to the temperature of the thermal plasma.

As noted in earlier analyses, the final relic density can be higher or
lower than in standard cosmology, depending on the values of various
free parameters. However, we find that even in this more generalized
scenario the density of DM particles with annihilation cross section
below that required of the usual thermal WIMPs will be too high,
unless the branching ratio of direct $\phi \rightarrow X'$ is below
$10^{-4} (M_{X^\prime}/100 \ {\rm GeV})$, {\em and} the $\phi$
particles are not too heavy,
$M_\phi \lsim 10^7 \ {\rm GeV} (M_{X^\prime}/100 \ {\rm GeV})^{2/3}$.
Recall that if the $\phi$ decay width is suppressed by
$M_{\rm Pl}^{-2}$, as in generic moduli or Polonyi models,
$M_\phi < 10^7$ GeV implies a rather low reheat temperature,
$T_{\rm RH} \lsim 1$ GeV. This bound can only be avoided if the
$X^\prime$ annihilation cross section is more than 10 orders of
magnitude below that of thermal WIMPs, in which case none of the usual
DM searches (direct, indirect and at colliders) is likely to yield a
signal. This bound has been derived under the assumption of vanishing
initial radiation and $X^\prime$ density. Since deviating from this
assumption can only increase the final DM density, it retains its
validity in full generality.

We also investigated quantitatively the impact of a non--vanishing
initial radiation density, parameterized by the ratio $\mu$ of initial
radiation and $\phi$ matter densities; we argued that in most cases
the initial density of DM particles is then also non--vanishing, and
can be estimated from the equilibrium density. The initial radiation
density is irrelevant if the DM annihilation cross section is so large
that DM particles attained thermal equilibrium during the period of
$\phi$ matter domination (and possibly thereafter) even for
$\mu=0$. On the other hand, for small DM annihilation cross section
even a small non--vanishing value of $\mu$ can have sizable effects,
unless the period of early matter domination is very long. We
parameterize this by the ratio $\gamma$ of the initial Hubble
parameter to the total $\phi$ decay width. We find that for very small
DM annihilation cross section the final DM density becomes independent
of $\mu$ only if
$\gamma > 10^{20} \mu^{3/2} \sqrt{1+\mu} (10^{-5}/B_{X^\prime})^2
(10^7 \ {\rm GeV} /M_\phi)$,
where $B_{X^\prime}$ is the branching ratio for direct
$\phi \rightarrow X^\prime$ decays. Moreover, there will only be an
extended period of $\phi$ matter domination if
$\gamma \gg \max(1,\, \mu^2)$ .

During the early period of $\phi$ matter domination
  density perturbations on scales smaller than the Hubble scale will
  grow linearly with the expansion parameter \cite{kt}. This enhances
  the perturbation spectrum at very small scales relative to standard
  cosmology.  However, even if the DM particles are produced
  non--thermally, in most cases we considered they will quickly attain
  kinetic equilibrium with the thermal plasma. This gives them a
  free--streaming length which is much bigger than the size of the
  density perturbations that get enhanced during the early matter
  domination, effectively erasing these perturbations again. The same
  conclusion holds for very weakly coupled DM particles that did not
  thermalize. They would have to be produced predominantly directly
  from $\phi$ decay. Unless the $X^\prime$ particles are already produced
  non--relativistically, e.g.  $M_{X^\prime} \simeq M_\phi/2$ for $2-$body
  $\phi \rightarrow X^\prime X^\prime$ decay, the free--streaming length of
  $X^\prime$ is too large for the early ``minihaloes'' to survive
  \cite{structure1,structure2}. Therefore the scenario considered in this paper in
  almost all cases reproduces the predictions of standard CDM as far
  as structure formation is concerned.

However, we saw above that an early epoch of matter
  domination allows to reproduce the correct DM relic density for a
  wide range of DM annihilation cross sections, which can be both
  smaller or larger than that required for thermal WIMPs in standard
  cosmology. The annihilation cross section can in principle be
  inferred by observing DM annihilation in today's universe (assuming
  the DM density at the point of annihilation is sufficiently well
  known). Moreover, the couplings of the DM particles can in principle
  be deduced from collider physics experiments \cite{colliders1,colliders2,colliders3}. One
  could then compute the annihilation cross section, and check whether
  it is compatible with the standard thermal WIMP scenario, or at
  least with the much larger range of cross sections that can be
  accommodated in our scenario. On the other hand, the $\phi$
  particles are so heavy, and so weakly coupled, that they will not be
  produced at colliders in the foreseeable future.

In this paper we assumed that the DM particle couples directly to SM
particles, allowing for the case of very weak couplings. In
ref.\cite{Kane:2015qea} a more complicated ``dark sector'' was
investigated, allowing for dark radiation (with temperature typically
smaller than that of the visible sector) and a WIMP--like parent
particle $X$ that can decay into $X^\prime$. While we did not perform
extensive numerical scans of this case, we note that an accurate
treatment of the temperature dependence of $h_*$ is as important in
this case as in the somewhat simpler case we considered. Moreover, we
expect the impact of a non--vanishing initial radiation density to be
comparable to that in our scenario, with the possible caveat that
assuming a very weakly coupled $X^\prime$ particle to be initially in
thermal equilibrium is probably less motivated than in the scenarios
we consider; however, the parent $X$ particle in the scenario of
ref.\cite{Kane:2015qea} should indeed initially have been in thermal
equilibrium.

Even in our somewhat simpler scenario we had to numerically track the
evolution of the $\phi$, radiation and DM densities over a very large
range of Hubble parameters, or time, which is computationally rather
expensive. In this paper we therefore used simple approximations for
the thermally averaged DM annihilation cross sections, in most cases
replacing it by a constant. In future publications we intend to
investigate specific well--motivated DM candidate particles in
scenarios with an early $\phi$ matter dominated epoch, including the
full energy (or temperature) dependence of the annihilation cross
section.

\acknowledgments
FH thanks the organizers of ``Post--Inflationary String Cosmology'' workshop in Bologna, September 2017, for their hospitality and support. 
This work was partially supported by
  the TR33 ``The Dark Universe'' funded by the Deutsche
  Forschungsgemeinschaft. FH was supported by the Deutsche Akademische Austauschdienst (DAAD).


\begin{thebibliography}{99}

\bibitem{Bertone:2016nfn} 
  G.~Bertone and D.~Hooper,
  [arXiv:1605.04909 [astro-ph.CO]].
  
\bibitem{Bertone:2004pz} 
  G.~Bertone, D.~Hooper and J.~Silk,
  Phys.\ Rept.\  {\bf 405}, 279 (2005), 
  doi:10.1016/j.physrep.2004.08.031
  [hep-ph/0404175].
  
\bibitem{Baer:2014eja} 
  H.~Baer, K.Y.~Choi, J.E.~Kim and L.~Roszkowski,
  Phys.\ Rept.\  {\bf 555}, 1 (2015),
  doi:10.1016/j.physrep.2014.10.002
  [arXiv:1407.0017 [hep-ph]].

\bibitem{susy}
G. Jungman, M. Kamionkowski and K. Griest, Phys. Rept. {\bf 267}, 195 (1996),
doi: 10.1016/0370-1573(95)00058-5 [hep-ph/9506380].

\bibitem{Angloher:2015ewa} 
  G.~Angloher {\it et al.} [CRESST Collaboration],
  Eur.\ Phys.\ J.\ {\bf C76}, 25 (2016),
  doi:10.1140/epjc/s10052-016-3877-3
  [arXiv:1509.01515 [astro-ph.CO]].
  
\bibitem{Agnese:2015nto}
  R.~Agnese {\it et al.} [SuperCDMS Collaboration],
  Phys.\ Rev.\ Lett.\  {\bf 116}, 071301 (2016),
  doi:10.1103/PhysRevLett.116.071301
  [arXiv:1509.02448 [astro-ph.CO]].
  
\bibitem{Akerib:2016vxi} 
  D.S.~Akerib {\it et al.} [LUX Collaboration],
  Phys.\ Rev.\ Lett.\  {\bf 118}, 021303 (2017),
  doi:10.1103/PhysRevLett.118.021303
  [arXiv:1608.07648 [astro-ph.CO]].
  
\bibitem{Tan:2016zwf} 
  A.~Tan {\it et al.} [PandaX-II Collaboration],
  Phys.\ Rev.\ Lett.\  {\bf 117}, 121303 (2016),
  doi:10.1103/PhysRevLett.117.121303
  [arXiv:1607.07400 [hep-ex]].
  
\bibitem{Ackermann:2015zua}
  M.~Ackermann {\it et al.} [Fermi-LAT Collaboration],
  Phys.\ Rev.\ Lett.\  {\bf 115}, 231301 (2015),
  doi:10.1103/PhysRevLett.115.231301
  [arXiv:1503.02641 [astro-ph.HE]].
  
\bibitem{Fermi-LAT:2016uux} 
  A.~Albert {\it et al.} [Fermi-LAT and DES Collaborations],
  Astrophys.\ J.\  {\bf 834}, 110 (2017),
  doi:10.3847/1538-4357/834/2/110
  [arXiv:1611.03184 [astro-ph.HE]].
  
\bibitem{Ahnen:2016qkx}
  M.L.~Ahnen {\it et al.} [MAGIC and Fermi-LAT Collaborations],
  JCAP {\bf 1602}, 039 (2016), 
  doi:10.1088/1475-7516/2016/02/039
  [arXiv:1601.06590 [astro-ph.HE]].

\bibitem{Easther:2013nga} 
R.~Easther, R.~Galvez, O.~Ozsoy and S.~Watson,
Phys.\ Rev.\ {\bf D89}, 023522 (2014),
doi:10.1103/PhysRevD.89.023522
[arXiv:1307.2453 [hep-ph]].

\bibitem{Arcadi:2011ev} 
  G.~Arcadi and P.~Ullio,
  Phys.\ Rev.\ {\bf D84}, 043520 (2011),
  doi:10.1103/PhysRevD.84.043520
  [arXiv:1104.3591 [hep-ph]].
  
\bibitem{polonyi}
J. Polonyi, Budapest preprint KFKI-1977-93, unpublished.

\bibitem{modprod1}
A. Vilenkin and L.H. Ford, Phys. Rev. {\bf D26}, 1231 (1982),
doi:10.1103/PhysRevD.26.1231.

\bibitem{modprod2}
A.D.~Linde, Phys. Lett. {\bf 116B}, 335 (1982),
doi:10.1016/0370-2693(82)90293-3.

\bibitem{modprod3}
A.A.~Starobinsky, Phys. Lett. {\bf 117B}, 175 (1982),
doi:10.1016/0370-2693(82)90541-X.

\bibitem{modprod4}
A.S.~Goncharov, A.D.~Linde and M.I.~Vysotsky, Phys.Lett. {\bf 147B}, 279
 (1984), doi: 10.1016/0370-2693(84)90116-3.
 
 \bibitem{modprod5}
M.~Dine, L.~Randall and S.D.~Thomas, Phys. Rev. Lett. {\bf 75}, 398 (1995),
doi: 10.1103/PhysRevLett.75.398 [hep-ph/9503303].

\bibitem{Coughlan:1983ci} 
G.D.~Coughlan, W.~Fischler, E.W.~Kolb, S.~Raby and G.G.~Ross,
Phys.\ Lett.\ {\bf B131}, 59 (1983),
doi:10.1016/0370-2693(83)91091-2

\bibitem{deCarlos:1993wie} 
B.~de Carlos, J.A.~Casas, F.~Quevedo and E.~Roulet,
Phys.\ Lett.\ {\bf B318}, 447 (1993),
doi:10.1016/0370-2693(93)91538-X
[hep-ph/9308325].

\bibitem{Banks:1993en} 
T.~Banks, D.B.~Kaplan and A.E.~Nelson,
Phys.\ Rev.\ {\bf D49}, 779 (1994),
doi:10.1103/PhysRevD.49.779
[hep-ph/9308292].

\bibitem{Ellis:1986zt} 
  J.R.~Ellis, D.V.~Nanopoulos and M.~Quiros,
  Phys.\ Lett.\ {\bf B174}, 176 (1986),
  doi:10.1016/0370-2693(86)90736-7
  
  
  \bibitem{Polnarev:1982} 
  A.~G.~Polnarev and M. Yu Khlopov, 
   Sov. Astron.(Engl. Transl.);(United States) 26.1 (1982).
  
  
\bibitem{Kawasaki:2000en} 
  M.~Kawasaki, K.~Kohri and N.~Sugiyama,
  Phys.\ Rev.\ D {\bf 62}, 023506 (2000)
  doi:10.1103/PhysRevD.62.023506
  [astro-ph/0002127].
  
\bibitem{Hannestad:2004px} 
S.~Hannestad,
Phys.\ Rev.\ {\bf D70}, 043506 (2004),
doi:10.1103/PhysRevD.70.043506
[astro-ph/0403291].

\bibitem{deSalas:2015glj} 
  P.F.~de Salas, M.~Lattanzi, G.~Mangano, G.~Miele, S.~Pastor and O.~Pisanti,
  Phys.\ Rev.\  {\bf D92}, 123534 (2015),
  doi:10.1103/PhysRevD.92.123534
  [arXiv:1511.00672 [astro-ph.CO]].
  
  
\bibitem{split1} 
  J.~D.~Wells,
  hep-ph/0306127.
  
\bibitem{split2}
G.F. Giudice and A. Romanino, Nucl. Phys. {\bf B699}, 65 (2004), Erratum:
Nucl. Phys. {\bf B706}, 487 (2005), 
doi:10.1016/j.nuclphysb.2004.11.048, 10.1016/j.nuclphys.2004.08.001
[hep-ph/0406088].

\bibitem{Acharya:2009zt} 
  B.S.~Acharya, G.L.~Kane, S.~Watson and P.~Kumar,
  Phys.\ Rev.\ {\bf D80}, 083529 (2009),
  doi:10.1103/PhysRevD.80.083529
  [arXiv:0908.2430 [astro-ph.CO]].
  
\bibitem{Gelmini:2006pw} 
  G.B.~Gelmini and P.~Gondolo,
  Phys.\ Rev.\ {\bf D74}, 023510 (2006),
  doi:10.1103/PhysRevD.74.023510
  [hep-ph/0602230].
  
\bibitem{Allahverdi:2002nb}
  R.~Allahverdi and M.~Drees,
  Phys.\ Rev.\ Lett.\  {\bf 89}, 091302 (2002),
  doi:10.1103/PhysRevLett.89.091302
  [hep-ph/0203118].

\bibitem{Pallis:2004yy} 
C.~Pallis,
Astropart.\ Phys.\  {\bf 21}, 689 (2004),
doi:10.1016/j.astropartphys.2004.05.006
[hep-ph/0402033].

\bibitem{Giudice:2000ex} 
  G.F.~Giudice, E.W.~Kolb and A.~Riotto,
  Phys.\ Rev.\ {\bf D64}, 023508 (2001),
  doi:10.1103/PhysRevD.64.023508
  [hep-ph/0005123].
  
\bibitem{Chung:1998rq} 
  D.J.H.~Chung, E.W.~Kolb and A.~Riotto,
  Phys.\ Rev.\ {\bf D60}, 063504 (1999),
  doi:10.1103/PhysRevD.60.063504
  [hep-ph/9809453].

\bibitem{Dev:2013yza} 
P.S.~Bhupal Dev, A.~Mazumdar and S.~Qutub,
Front.\ in Phys.\  {\bf 2}, 26 (2014),
doi:10.3389/fphy.2014.00026
[arXiv:1311.5297 [hep-ph]].
    
\bibitem{Allahverdi:2013noa} 
  R.~Allahverdi, M.~Cicoli, B.~Dutta and K.~Sinha,
  Phys.\ Rev.\ {\bf D88}, 095015 (2013),
  doi:10.1103/PhysRevD.88.095015
  [arXiv:1307.5086 [hep-ph]].
  
\bibitem{Nakamura:2006uc} 
S.~Nakamura and M.~Yamaguchi,
Phys.\ Lett.\ {\bf B638}, 389 (2006),
doi:10.1016/j.physletb.2006.05.078
[hep-ph/0602081].

\bibitem{Asaka:2006bv} 
T.~Asaka, S.~Nakamura and M.~Yamaguchi,
Phys.\ Rev.\ {\bf D74}, 023520 (2006),
doi:10.1103/PhysRevD.74.023520
[hep-ph/0604132].

\bibitem{Feng:2004mt}
  J.L.~Feng, S.~Su and F.~Takayama,
  Phys.\ Rev.\ {\bf D70} (2004) 075019,
  doi:10.1103/PhysRevD.70.075019
  [hep-ph/0404231].

\bibitem{Endo:2006zj} 
  M.~Endo, K.~Hamaguchi and F.~Takahashi,
  Phys.\ Rev.\ Lett.\  {\bf 96}, 211301 (2006),
  doi:10.1103/PhysRevLett.96.211301
  [hep-ph/0602061].
 
\bibitem{Blumenhagen:2009gk} 
  R.~Blumenhagen, J.P.~Conlon, S.~Krippendorf, S.~Moster and F.~Quevedo,
  JHEP {\bf 0909}, 007 (2009),
  doi:10.1088/1126-6708/2009/09/007
  [arXiv:0906.3297 [hep-th]].
  
\bibitem{Aparicio:2014wxa} 
  L.~Aparicio, M.~Cicoli, S.~Krippendorf, A.~Maharana, F.~Muia and F.~Quevedo,
  JHEP {\bf 1411}, 071 (2014),
  doi:10.1007/JHEP11(2014)071
  [arXiv:1409.1931 [hep-th]].
  
\bibitem{Hall:2009bx} 
  L.J.~Hall, K.~Jedamzik, J.~March-Russell and S.M.~West,
  JHEP {\bf 1003}, 080 (2010),
  doi:10.1007/JHEP03(2010)080
  [arXiv:0911.1120 [hep-ph]].

\bibitem{Kane:2015qea} 
  G.L.~Kane, P.~Kumar, B.D.~Nelson and B.~Zheng,
  Phys.\ Rev.\ {\bf D93}, 063527 (2016),
  doi:10.1103/PhysRevD.93.063527
  [arXiv:1502.05406 [hep-ph]].

\bibitem{Ade:2015xua} 
  P.A.R.~Ade {\it et al.} [Planck Collaboration],
  Astron.\ Astrophys.\  {\bf 594}, A13 (2016),
  doi:10.1051/0004-6361/201525830
  [arXiv:1502.01589 [astro-ph.CO]].
  
\bibitem{Drees:2015exa} 
  M.~Drees, F.~Hajkarim and E.R.~Schmitz,
  JCAP {\bf 1506}, 025 (2015),
  doi:10.1088/1475-7516/2015/06/025
  [arXiv:1503.03513 [hep-ph]].
  
\bibitem{Kajantie:1996mn} 
  K.~Kajantie, M.~Laine, K.~Rummukainen and M.E.~Shaposhnikov,
  Phys.\ Rev.\ Lett.\  {\bf 77}, 2887 (1996),
  doi:10.1103/PhysRevLett.77.2887
  [hep-ph/9605288].
  
\bibitem{Csikor:1998eu} 
  F.~Csikor, Z.~Fodor and J.~Heitger,
  Phys.\ Rev.\ Lett.\  {\bf 82}, 21 (1999),
  doi:10.1103/PhysRevLett.82.21
  [hep-ph/9809291].
  
\bibitem{Fodor:1999at} 
  Z.~Fodor,
  Nucl.\ Phys.\ Proc.\ Suppl.\  {\bf 83}, 121 (2000),
  doi:10.1016/S0920-5632(00)91603-7
  [hep-lat/9909162].
    
\bibitem{Bazavov:2014pvz} 
  A.~Bazavov {\it et al.} [HotQCD Collaboration],
  Phys.\ Rev.\ {\bf D90}, 094503 (2014),
  doi:10.1103/PhysRevD.90.094503
  [arXiv:1407.6387 [hep-lat]].
  
\bibitem{Huovinen:2009yb} 
  P.~Huovinen and P.~Petreczky,
  Nucl.\ Phys.\ {\bf A837}, 26 (2010),
  doi:10.1016/j.nuclphysa.2010.02.015
  [arXiv:0912.2541 [hep-ph]].
  
\bibitem{Lesgourgues:2012uu} 
  J.~Lesgourgues and S.~Pastor,
  Adv.\ High Energy Phys.\  {\bf 2012}, 608515 (2012),
  doi:10.1155/2012/608515
  [arXiv:1212.6154 [hep-ph]].

\bibitem{kt}
E.W. Kolb and M.S. Turner, ``The Early Universe'', Front.Phys. 69 (1990).

\bibitem{Kane:2015jia} 
  G.L.~Kane, K.~Sinha and S.~Watson,
  Int.\ J.\ Mod.\ Phys.\ {\bf D24}, 1530022 (2015),
  doi:10.1142/S0218271815300220
  [arXiv:1502.07746 [hep-th]].

\bibitem{allah2}
R. Allahverdi and M. Drees, Phys. Rev. {\bf D66}, 063513 (2002),
doi: 10.1103/PhysRevD.66.063513 [hep-ph/0205246].

\bibitem{Harigaya:2013vwa} 
  K.~Harigaya and K.~Mukaida,
  JHEP {\bf 1405}, 006 (2014)
  doi:10.1007/JHEP05(2014)006
  [arXiv:1312.3097 [hep-ph]].

\bibitem{Harigaya:2014waa} 
  K.~Harigaya, M.~Kawasaki, K.~Mukaida and M.~Yamada,
  Phys.\ Rev.\ D {\bf 89}, no. 8, 083532 (2014)
  doi:10.1103/PhysRevD.89.083532
  [arXiv:1402.2846 [hep-ph]].
  
\bibitem{Mukaida:2015ria} 
  K.~Mukaida and M.~Yamada,
  JCAP {\bf 1602}, no. 02, 003 (2016)
  doi:10.1088/1475-7516/2016/02/003
  [arXiv:1506.07661 [hep-ph]].
  
\bibitem{Agashe:2014KDa} 
  K.A.~Olive {\it et al.} [Particle Data Group],
  Chin.\ Phys.\ {\bf C38}, 090001 (2014),
  doi:10.1088/1674-1137/38/9/090001
  
\bibitem{Hamdan:2017psw} 
  S.~Hamdan and J.~Unwin,
  arXiv:1710.03758 [hep-ph].

  
\bibitem{dik}
M. Drees, H. Iminniyaz and M. Kakizaki, Phys. Rev. {\bf D73}, 123502 (2006),
doi: 10.1103/PhysRevD.73.123502 [hep-ph/0603165].

\bibitem{moroi}
T. Moroi and L. Randall, Nucl. Phys. {\bf B570}, 455 (2000),
doi:10.1016/S0550-3213(99)00748-8 [hep-ph/9906527].
  
\bibitem{Co:2015pka} 
  R.T.~Co, F.~D'Eramo, L.J.~Hall and D.~Pappadopulo,
  JCAP {\bf 1512}, 024 (2015),
  doi:10.1088/1475-7516/2015/12/024
  [arXiv:1506.07532 [hep-ph]].
  
\bibitem{Linde:1983gd} 
  A.D.~Linde,
  Phys.\ Lett.\  {\bf 129B}, 177 (1983),
  doi:10.1016/0370-2693(83)90837-7
  
  \bibitem{structure1}
  A.~L.~Erickcek and K.~Sigurdson,
  Phys.\ Rev.\ D {\bf 84}, 083503 (2011)
  doi:10.1103/PhysRevD.84.083503
  [arXiv:1106.0536 [astro-ph.CO]].
  
  \bibitem{structure2}
  J. Fan, O. \"Ozsoy and S. Watson, Phys. Rev. {\bf D90}, 043536 (2014),
doi: 10.1103/PhysRevD.90.043536 [arXiv:1405.7373 [hep-ph]].


\bibitem{colliders1}
M. Brhlik, D.J.H. Chung and G.L. Kane,
Int. J. Mod. Phys. {\bf D10}, 367 (2001),
doi: 10.1142/S0218271801000998 [hep-ph/0005158].
\bibitem{colliders2}
M. Drees, Y.G. Kim, M.M. Nojiri, D. Toya, K. Hasuko and T. Kobayashi,
Phys. Rev. {\bf D63}, 035008 (2001),
doi: 10.1103/PhysRevD.63  [hep-ph/0007202].
\bibitem{colliders3}
V.A. Mitsou, Int. J. Mod. Phys. {\bf A28}, 1330052 (2013),
doi: 10.1142/S0217751X13300524 [arXiv:1310.1072 [hep-ex]].
  


\end{thebibliography}
\end{document}